\documentclass[prl,aps,twocolumn,superscriptaddress]{revtex4-2}
\usepackage{graphicx,color,setspace}
\usepackage[utf8]{inputenc}
\usepackage{mathtools}
\usepackage[normalem]{ulem}
\usepackage{tabularx}
\usepackage{enumerate}

\begin{document}
\title{A simple solution to the problem of self-assembling cubic diamond crystals}
\author{Lorenzo Rovigatti}
\affiliation{Dipartimento di Fisica, Sapienza Universit\`{a} di Roma, P.le Aldo Moro 5, 00185 Rome, Italy}
\affiliation{CNR-ISC Uos Sapienza, Piazzale A. Moro 2, IT-00185 Roma, Italy}
\author{John Russo}
\affiliation{Dipartimento di Fisica, Sapienza Universit\`{a} di Roma, P.le Aldo Moro 5, 00185 Rome, Italy}
\author{Flavio Romano}
\affiliation{Dipartimento di Scienze Molecolari e Nanosistemi, Universit\`{a} Ca' Foscari di Venezia Campus Scientifico, Edificio Alfa, via Torino 155, 30170 Venezia Mestre, Italy}
\affiliation{European Centre for Living Technology (ECLT) Ca' Bottacin, 3911 Dorsoduro Calle Crosera, 30123 Venice, Italy}
\author{Michael Matthies}
\affiliation{School of Molecular Sciences and Center for Molecular Design and Biomimetics, The Biodesign Institute, Arizona State University, 1001 South McAllister Avenue, Tempe, Arizona 85281, USA}
\author{Luk\'{a}\v{s} Kroc}
\affiliation{School of Molecular Sciences and Center for Molecular Design and Biomimetics, The Biodesign Institute, Arizona State University, 1001 South McAllister Avenue, Tempe, Arizona 85281, USA}
\author{Petr \v{S}ulc}
\affiliation{School of Molecular Sciences and Center for Molecular Design and Biomimetics, The Biodesign Institute, Arizona State University, 1001 South McAllister Avenue, Tempe, Arizona 85281, USA}


\begin{abstract}
The self-assembly of colloidal diamond (CD) crystals is considered as one of the most coveted goals of nanotechnology, both from the technological and fundamental points of view. For applications, colloidal diamond is a photonic crystal which can open new possibilities of manipulating light for information processing. From a fundamental point of view, its unique symmetry exacerbates a series of problems that are commonly faced during the self-assembly of target structures, such as the presence of kinetic traps and the formation of crystalline defects and alternative structures (polymorphs). Here we demonstrate that all these problems can be systematically addressed via SAT-assembly, a design framework that converts self-assembly into a satisfiability problem. Contrary to previous solutions (requiring four or more components), we prove that the assembly of the CD crystal only requires a binary mixture. Moreover, we use molecular dynamics simulations of a system composed by nearly a million nucleotides to test a DNA nanotechnology design that constitutes a promising candidate for experimental realization.
\end{abstract}

\maketitle

The bottom-up realization of complex structures from elementary components has been the main goal of nanotechnology since Feynman first imagined "What would happen if we could arrange the atoms one by one the way we want them"~\cite{feynman2018there}. In the last decade, one of the most sought-after arrangements is that of a cubic diamond (CD) crystal. Realizing a diamond structure on the colloidal scale would in fact open the doors to the realization of optical metamaterials~\cite{soukoulis2010optical,ngo2006tetrastack}. In analogy to semiconductors, which have an electronic band gap, a metamaterial lattice can be designed to possess an optical band gap, which prevents the transport of photons within a specific range of wavelengths through the material. If such a lattice can be assembled with cavity sizes comparable to the wavelength of visible or infrared light, it would open new possibilities of manipulating such radiation in the same way semiconductors manipulate electrons.

Efforts to reliably produce diamond structures have spanned many decades of research and have employed a variety of techniques, from traditional top-down approaches like lithography and holography~\cite{werts2002nanometer}, to bottom-up strategies like the self-assembly of liquid crystals, nanoparticles and colloidal particles~\cite{ducrot2017colloidal,liu2016diamond}. To reach the lengthscales that are required for photonic materials, self-assembly has emerged as one of the most promising routes for the realization of diamond cubic structure. Notwithstanding recent successes \cite{liu2016diamond,michelson2022three,he2020colloidal}, realizing a diamond lattice remains notoriously difficult, and the reasons are manyfold.

\begin{figure}[!ht]
    \centering
    \includegraphics[width=.5\textwidth]{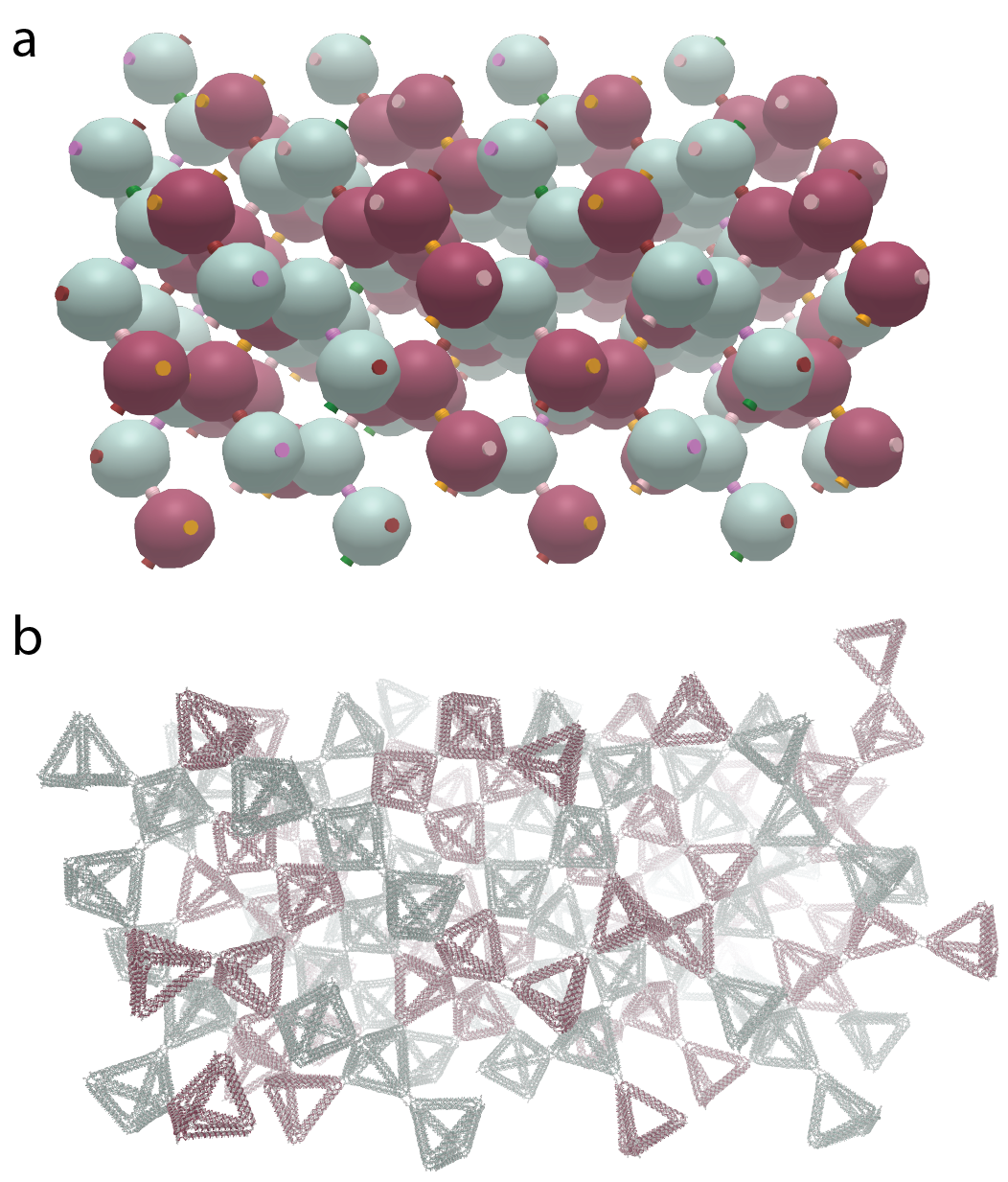}
    \caption{
      Proposed realization of the CD lattice crystal. \textbf{a} Finite size cluster of 128 Patchy Particles (the 16-particle cubic diamond unit cell is multiplied 2x2x2), which interact anisotropically through colored patches according to the  interaction matrix provided by SAT-assembly. \textbf{b} Mean structure of the CD lattice assembled from two distinct types of DNA origamis, as represented by the oxDNA model.
    }
    \label{fig_experimental}
\end{figure}

The first major problem is represented by the low volume fraction of the CD crystal. The CD crystal is an open crystalline structure which is mechanically stable at volume fractions that are almost half those of ordinary closed packed structures, like fcc and bcc. This makes the assembly particularly difficult for the isotropic and short-range potentials that typically govern colloidal particles, and which tend to favour close-packed structures. To overcome this problem, one of the most promising approaches has been to use anisotropic interactions~\cite{russo2021physics}, and in particular colloidal particles with tetrahedrally arranged bonding sites. These particles are termed \emph{Patchy Particles} (Fig.~\ref{fig_experimental}a) and they acquire anisotropic interactions either via their shape~\cite{van2013entropically} or via chemical patterning of their surface~\cite{zhang2004self,pawar2010fabrication,bianchi2011patchy,romano2011colloidal,suzuki2009controlling,kim2011dna,wang2012colloids,feng2013dna}.

The second common problem with the assembly process is that during the nucleation process different crystalline phases (called \emph{polymorphs}) can nucleate~\cite{mahynski2016bottom,leoni2021non}. Systems that can assemble a CD lattice are often found to be able to assemble into the hexagonal diamond (HD) lattice, resulting in imperfect crystals with defects and stacking faults. Avoiding or minimizing the formation of polymorphs requires specific solutions: to avoid the formation of stacking faults in CD formations usually requires either torsional interactions~\cite{romano2012patterning,tracey2019programming} or hierarchical assembly~\cite{morphew2018programming,patra2018programmable,tracey2019programming,ma2019inverse}. Importantly, some of these solutions lack a convincing experimental counterpart.

The final obstacle to a successful self-assembly strategy is the presence of \emph{kinetic traps}. Despite the reversible nature of the bonds between different particles, the thermodynamic equilibrium of self-assembling systems can be easily derailed by locally stable minima that force the system to stay in a kinetically trapped state. Multiple strategies have emerged to avoid kinetic traps, from suppressing local motifs found in kinetic traps but not in the desired structures (such as odd-numbered rings~\cite{neophytou2021facile}) to designing specific protocols that steer the non-equilibrium assembly in the correct direction~\cite{bisker2018nonequilibrium,whitelam2020learning,whitelam2021neuroevolutionary,bupathy2021temperature}.

Recently we have introduced a novel framework, named SAT-assembly~\cite{romano2020designing,russo2021sat}, to design patchy particle systems that can assemble into any desired structure. Through a mapping to a Boolean satisfiability problem (SAT), the structure is encoded into the interaction map describing the bonds that can be formed between the different particles. Crucially, the interaction map can also encode the avoidance of alternative structures, which can be both competing polymorphs or commonly occurring kinetic traps. We can then use highly optimized SAT-solver algorithms to find solution in terms of interaction matrix between patchy particles that assemble target lattice and avoid specified alternative structures. Here we systematically explore the space of solutions that can form the CD crystal while avoiding the HD polymorph, and demonstrate that the minimum number of particle species required for the defect-free assembly of CD is only 2. We thus present the binary mixture solution that encodes this design, and propose a possible realization as a DNA origami design, which we study with oxDNA, a coarse-grained model of DNA \cite{snodin2016direct,rovigatti2015comparison,ouldridge2011structural,sulc2012sequence} (Fig.~\ref{fig_experimental}b).
Via molecular simulations, we explore the nucleation pathway of the solution and reveal that it occurs through a two-step nucleation process aided by gas-liquid demixing. Requiring the minimal number of components (a binary mixture), together with its facile crystallizability, the solution is the most promising candidate for experimental realization of a colloidal diamond crystal yet using patchy particles without torsional modulation of their interacting patches, which allows for realization e.g. with gold nanoparticles or colloids with patches realized as sptially localized DNA brushes \cite{xiong2020three} or with DNA origamis with patches as single-stranded overhangs \cite{tian2020ordered}.


\section{The minimal solution}

\begin{table}[!t]
\centering
\begin{tabular}{|l|llll|}
\hline
PP species & \multicolumn{4}{|c|}{Patch Coloring}\\
\hline
\mbox{1: }  &(1,A) & (2,B) & (3,C) & (4,D) \\
\mbox{2: }  &(1,E) & (2,F) & (3,G) & (4,H) \\
\hline
\multicolumn{5}{|c|}{Color interactions}\\
\hline
\multicolumn{5}{|c|}{(A,H),(B,B),(C,C),(D,F),(E,G)} \\
\hline
\end{tabular}
\caption{Designed patchy particles for self-assembly into a cubic diamond crystal lattice. It consists of 2 patchy particle species and 8 colors. The patch coloring is in format (patch number, patch color), where patches are numbered from 1 to 4, and colors are assigned a letter from A to H. The color interaction then lists all pairs of interacting colors, where colors B and C are self-complementary. \label{table:four}}
\label{tab_ts}
\end{table}

Since it is desirable to keep the number of distinct particle types as low as possible for ease of manufacturing and experimental preparation, we seek to find the minimum set of particle species that can assemble into a CD lattice without any defects or alternative assemblies.
We formulate the design problem as a SAT problem~\cite{romano2020designing,russo2021sat}, which is a collection of Boolean clauses (see Methods and Supp. Table~S1) that encode the topology as given by the 16-particle unit cell for CD lattice (Supp.~Table S2). We then gradually enumerate all possible solutions by augmenting the set of conditions by additional clauses that contain negations of
already discovered solutions, so that the SAT problem finds new previously unknown solutions. We then test each discovered solution for its ability to assemble the hexagonal diamond lattice with 32 sites in the unit cell (Supp. Table~S3).


The SAT procedure proves that it is not possible to obtain a solution that would avoid hexagonal diamond formation with a single particle type ($N_s = 1$ and $1 \leq N_c \leq 4$ ). Therefore, we next scanned all patchy particle systems with $N_s = 2$ and $N_c = 8$ and found that there exists a solution to the SAT-assembly problem that is able to form a CD cubic lattice $16$-particle unit cell and avoids at the same time formation of the HD lattice with $32$-particle unit cell. The solution is listed in Table \ref{tab_ts}. Each of the two particle types have the same stoichiometry in the CD lattice, so for the full crystal assembly they are mixed in equal concentrations.

\section{DNA Lattice Simulation}

\begin{figure}
    \centering
    \includegraphics[width=.49\textwidth]{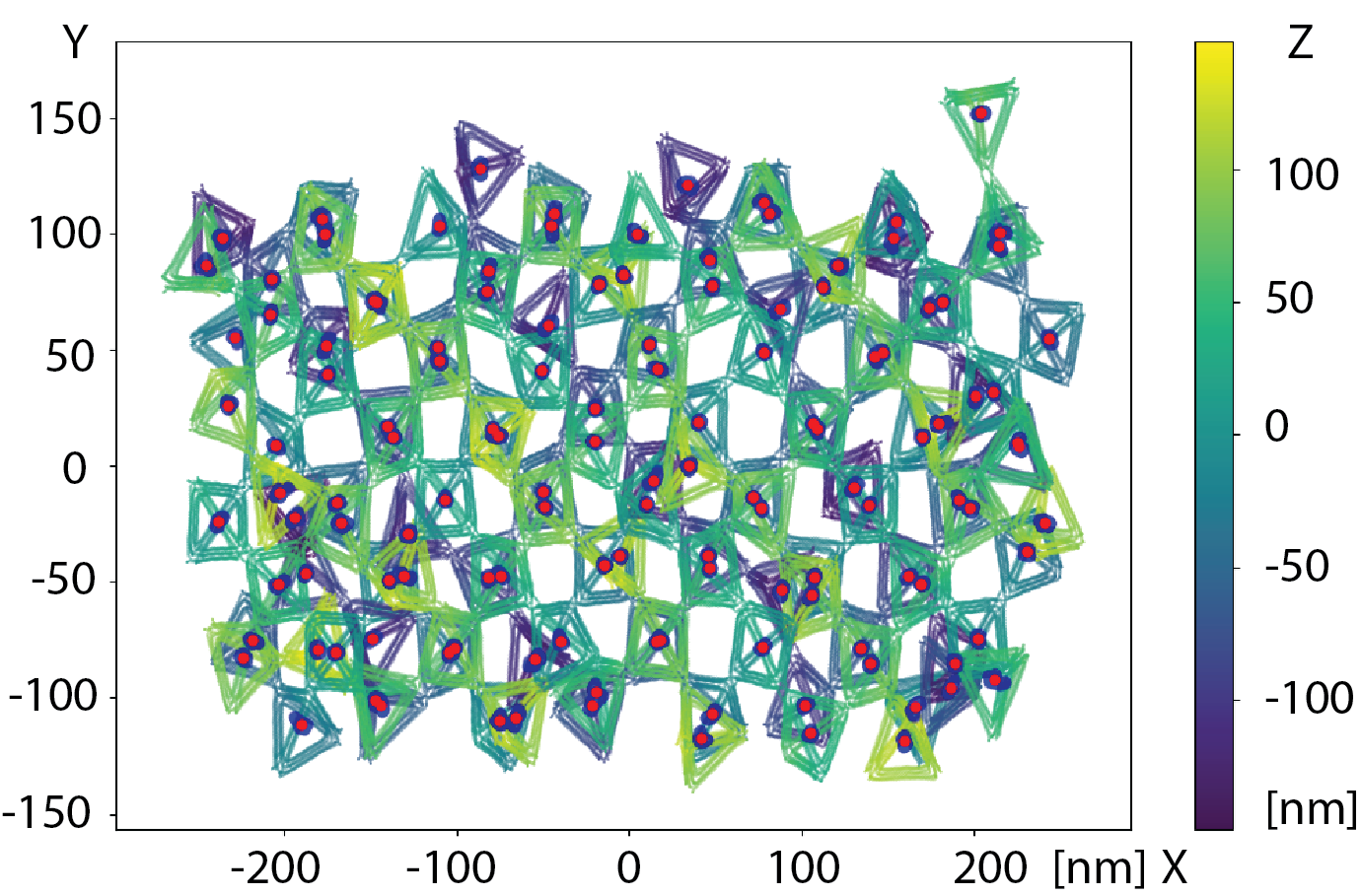}
    \caption{
       Projection of the mean structure of Fig.\ref{fig_experimental}b on the x-y plane. The z coordinate of each particle is encoded in the coloring. Red markers represent the centre of mass of every DNA origami particle computed using the mean structure. The blue point cloud around the red markers corresponds to the centre of mass positions of the structures during the production run, aligned to the centre of mass position of the mean structure (red markers).    
    }
    \label{fig_meanl}
\end{figure}

We start by mapping the minimal solution in the oxDNA model to simulate a proposed tetrahedral design realization (Fig.~\ref{fig_experimental}b). The DNA tetrahedron has single-stranded overhangs in each of its vertices that correspond to the DNA patches. As strands in each vertex are identical, they do not impose any torsional restraint in terms of orientations at which two particles can bind to each other, as opposed to e.g. shape complementary DNA origami designs~\cite{gerling2015dynamic}, which presents larger experimental challenge to assemble and correctly optimize the structure design to ensure pairing of the shape complementary regions \cite{park2019design}. 
Patches of complementary colors would correspond to Watson-Crick base pairing complementary single-stranded DNA overhangs, and self-complementary color would correspond to a palindromic DNA sequence. After assembling the CD lattice cluster out of individual DNA origamis
as described in Methods, we ran a sampling molecular dynamics simulations at $T = 293$ K. The system consists of 951040 bases, making it the largest nucleotide-level coarse-grained DNA simulation that has been realized to date.
The mean structure is the assembled cluster obtained from the MD simulation and it is shown in Fig.~\ref{fig_experimental}b. Its projection in Fig.~\ref{fig_meanl} shows how the centre of mass of each DNA origami moves relative to its mean structure throughout the simulation.
Our oxDNA model shows that the target lattice is an energy minima of the proposed design (\textit{i.e.} it is mechanically stable), and is a candidate for experimental realization.

We note that despite oxDNA model being highly coarse-grained, representing each nucleotide as a single rigid body, sampling the kinetics of assembly of individual origamis into the target lattice is still out of reach.
In the following section we address these limitations by adopting a more efficient (and more coarse-grained) Patchy Particle model.

\section{Patchy Particle Simulation Results}

\begin{figure*}[!ht]
    \centering
    \includegraphics[width=0.95\textwidth]{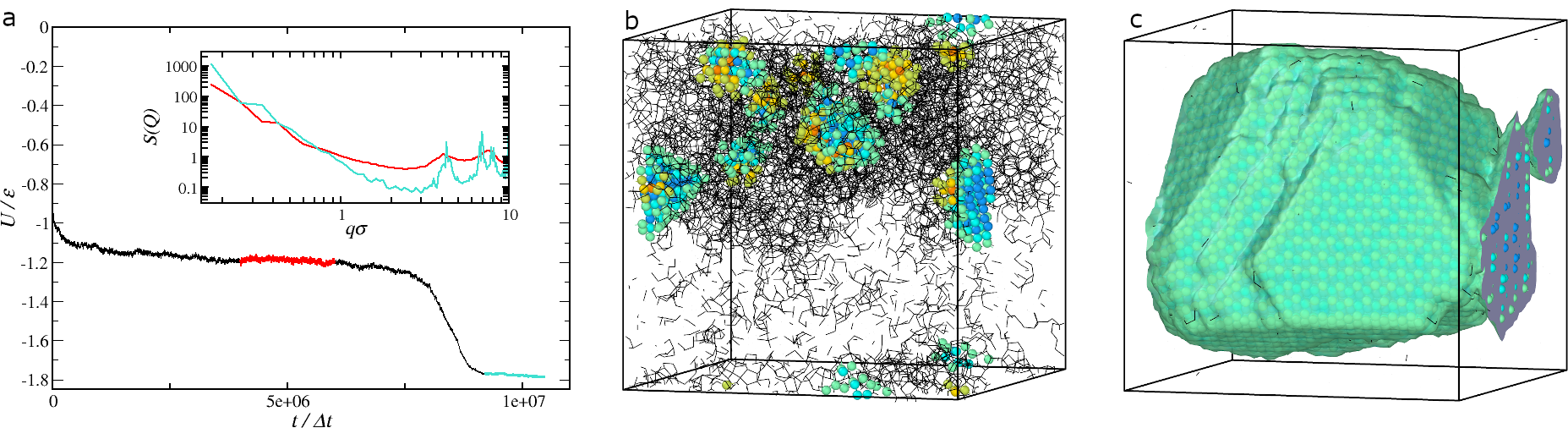}
    \caption{Nucleation run at $k_BT/\epsilon=0.095$ and $\rho\sigma^3=0.2$. \textbf{a} Potential energy as a function of simulation time steps. Parts of the trajectory are highlighted in red for the metastable liquid phase and in cyan for the crystalline phase. (inset) Structure factor corresponding to the highlighted parts of the trajectory. $\mathbf{b}$ Snapshot of a configuration in the metastable liquid phase. Particles in crystalline environments are depicted in blue for the CD crystal, and orange for the HD crystal. The color shade represents the distance from the surface. Particles in disordered environments are hidden and only bonds between them are represented with black lines. $\mathbf{c}$ Snapshot of a configuration in the crystal phase.
    A single CD crystal is formed, for which we plot a surface mesh to highlight the growth of crystalline planes.
    }
    \label{fig_nucleation}
\end{figure*}

To demonstrate the correct assembly of a CD crystal from our design, in Fig.~\ref{fig_nucleation} we show results from molecular dynamics simulations at temperature $k_BT/\epsilon=0.095$ and number density $\rho\sigma^3=0.2$ for $N = 10000$ particles using a generalization of the Kern-Frenkel Patchy Particle model with differentiable interaction potentials (dKF model, introduced in Methods). Fig.~\ref{fig_nucleation}a plots the potential energy as a function of the simulated time for a significant trajectory. The trajectory quickly reaches a transient state (partially colored in red), where the energy relaxation is very slow, followed by an abrupt change towards a low-energy state (partially colored in cyan). A comparison of the structure factor $S(q)$ for these two states (inset in Fig.~\ref{fig_nucleation}a) shows that the first one is characterized by an amorphous structure at high q-vectors (short distance), but a large inhomogeneity at small q-vectors (with a peak appearing at $q\sigma \to 0$), characteristic of a phase-separated system. The low-energy state shows a similar density inhomogeneity, but it is also characterized by sharp  peaks at finite $q$, typical of a crystalline state. These observations are confirmed by a visual inspection of these states (Fig.~\ref{fig_nucleation} b and c). Fig.~\ref{fig_nucleation}b is a snapshot taken in the phase-separated state. Particles in crystalline environments are depicted in blue for the CD crystal, and in orange for the HD crystal. Particles in disordered environments are represented with bond lines. The snapshot shows that the system has phase-separated into a liquid phase (top part of the simulation box) and a gas phase (bottom half of the simulation box). Inside the liquid phase there is an equilibrium population of pre-critical crystalline clusters that constantly form and dissolve. Note that for these small clusters the hexagonal symmetry (orange particles) is still present, albeit only on the surface of the nuclei. Fig.~\ref{fig_nucleation}c shows the final crystalline state into which the system self-assembles. A single CD crystal is formed, and an enveloping surface is drawn to highlight the different crystalline planes. The snapshot shows the lateral growth of the (1,1,1) plane.

\begin{figure}[!t]
    \centering
    \includegraphics[width=0.5\textwidth]{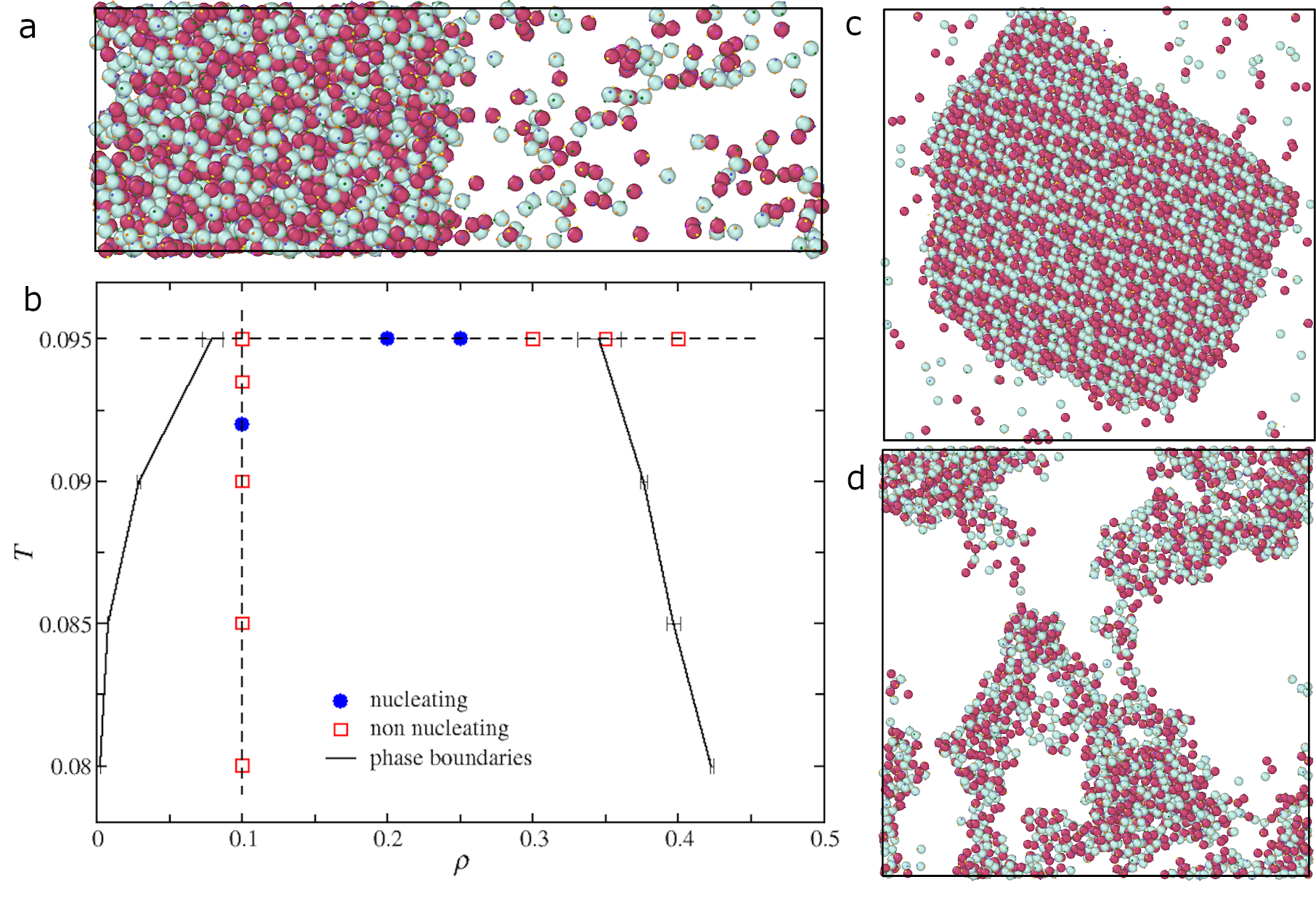}
    \caption{\textbf{a}-\textbf{b} Isoplethic phase diagram at equimolar concentration obtained via direct coexistence simulations (see snapshot \textbf{a}). Black lines are the liquid-gas coexistence densities. Error bars represent the dispersion over three independent trajectories at total density $\rho \sigma^3=0.5,0.55,0.6$. The horizontal dashed line connect simulations runs at $k_B T / \epsilon=0.095$ starting from homogeneous conditions, while the vertical dashed line connects the points at $\rho \sigma^3=0.1$. Full (blue) symbols represent trajectories that crystalline spontaneously, while open (red) symbols represent states that remain in a disordered phase. Slabs of width $10\sigma$ for $\rho \sigma^3 =0.1$ and \textbf{c} $k_B T / \epsilon =0.092$ and \textbf{d} $k_B T / \epsilon=0.08$.
    }
    \label{fig_coexistence}
\end{figure}

The nucleation pathway described so far is akin to a two-step nucleation process in which nucleation is preceded by the condensation of dense liquid droplets. To better analyse the nucleation pathway we map the gas-liquid phase diagram of the system via direct-coexistence simulations. The system is first equilibrated in the liquid state at a density $\rho$ and then put in an elongated box ($L_x=3L_y=3L_z$), where the slab of liquid is bounded by two interfaces with the gas parallel to the $yz$ plane. Simulations are then run at constant density and constant temperature until liquid-gas equilibrium is reached. A snapshot from an equilibrated configuration is shown in Fig.~\ref{fig_coexistence}a. We run direct coexistence simulations at densities $\rho \sigma^3=0.5,0.55,0.6$ and temperature ranging from $k_B T / \epsilon=0.08$ and $k_B T / \epsilon=0.1$. At the highest temperature, $k_B T / \epsilon=0.1$, the final equilibrium state consist of a single fluid phase (supercritical state) while the other temperatures maintain the liquid-slab configuration. From these we can measure equilibrium concentration and densities of the two phases. In Fig.~\ref{fig_coexistence}b we plot the isoplethic phase diagram at constant concentration $x=0.5$ between the two species: the black line indicates the coexistence density of the gas (left black branch) and liquid (right black branch) phases. In order to nucleate, the system is quenched inside the coexistence region, where the formation of liquid droplets favours the nucleation process. This is tested with several independent simulations of $N=10000$ patchy particles prepared in a homogeneous state and then quenched inside at $k_B T / \epsilon=0.095$ at different densities. Each simulation is represented as a point in Fig.~\ref{fig_coexistence}b joined by the horizontal dashed line. Open red symbols represent simulations which did not crystallize within the simulation time, while blue full symbols represent systems that spontaneously nucleated. Fig.~\ref{fig_coexistence}b shows that the nucleation rate has a maximum inside the coexistence region, located approximately under the critical point.
We also quench several systems at $\rho \sigma^3=0.1$ at different temperatures (joined by the vertical dashed line). These simulation show that the nucleation rate also has a maximum in temperature. A snapshot of a slab of the system at ($\rho \sigma^3=0.1$, $k_B T / \epsilon=0.092$) is shown in Fig.~\ref{fig_coexistence}c, which highlights the typical formation of a cubic crystal. While lowering the temperature guarantees a smaller nucleation barrier for nucleation, at low temperature ($k_B T / \epsilon \leq 0.09$) the dynamics of the system becomes arrested and instead of compact liquid droplets the system is frozen in a gel-state. A snapshot of this state at ($\rho \sigma^3=0.1$, $k_B T / \epsilon=0.08$) is shown in Fig.~\ref{fig_coexistence}d.

\section{Discussion and Conclusions}
We have used an inverse design framework based on formulating the design problem as a Boolean Satisfiability Problem to obtain the minimum set of patchy particles that, without torsional potential, can assemble into a Cubic Diamond (CD) lattice. We have demonstrated that at least two different types of patchy particles are required to obtain a system that self-assembles into the desired lattice and avoid the competing hexagonal diamond geometry, and have provided an explicit solution with two species and eight colours. We have observed that in the patchy particle simulations the system first forms a metastable liquid phase, from which the cubic diamond then crystallizes. Finally, we have showed a possible realization of the CD lattice made out of DNA tetrahedrons designed and verified using tools for DNA nanotechnology and coarse-grained simulations.

Our methods can also be used to design a system that avoids certain kinetic intermediates, thus possibly using SAT-assembly approach to optimize not only the yield of the target structure, but also the assembly kinetics. The rearrangement from a metastable liquid phase into the crystal lattice is a slow process, and our future work will explore if the SAT solver can be used to design interactions that further destabilize the liquid phase, thus enhancing the rate of formation of crystal lattice, as well as the effects of changing the number of different species and distinct patch colors onto the overall phase diagram of the system.

Our inverse design approach can be further generalized to other geometries, including finite-size 2D or 3D multicomponent nanostructures, with applications to biotemplated self-assembly and optical metamaterial manufacturing.

A more efficient model would be needed to capture the assembly kinetics of the DNA tetrahedrons, such as an intermediate multiscale model between Patchy Particle and oxDNA representation, which might represent part of the structure at nucleotide level (such as ssDNA overhangs) and part of the structure as a rigid body. Studies and development of such models will also be the focus of future work.

\section{Acknowledgments}
JR acknowledges support from the European Research Council Grant DLV-759187. P\v{S} acknowledges support from the ONR Grant N000142012094. JR and P\v{S} acknowledge support from the Universit{\`a} Ca' Foscari for a Visiting Scholarship and from NSF 1931487 - ERC DLV-759187 research collaboration grant. P\v{S} further acknowledges the use of the Extreme Science and Engineering Discovery Environment (XSEDE), which is supported by National Science Foundation grant number TG-BIO210009.


\section{Methods}
\subsection{SAT-assembly}
We use the SAT-assembly framework to find the colorings of patches of the patchy particles which are compatible with the desired target structure. The procedure has been described in detail in Refs.~\cite{russo2021sat,romano2020designing}. Briefly, we formulate the inverse design problem as a Boolean Satisfiability Problem. The parameters of the SAT-assembly are the number of species of patchy particles ($N_s = 2$ in our case), and the number of colors $N_c$ that the patchy particles will use. The interaction between particles is then determined by the color interaction matrix, which sets which patch colors can bind to each other (also allowing for self-complementarity). In the case of realization of patchy particles with DNA nanostructures such as DNA origami \cite{Rothemund2006,tian2020ordered}, complementary coloring of patches would corresponds to \textit{e.g.} a single-stranded DNA overhangs with complementary sequences that can hybridize into a duplex. Self-complementary colors would then correspond to palindromic DNA sequences.

To formulate the Boolean Satisfiability problem, each possible coloring and color interaction is assigned a Boolean variable which can be either true or false. Furthermore, we define Boolean variables that correspond to which positions in the target CD lattice are occupied by what type of particle, and how each respective patchy particle needs to be placed in each of the lattice positions as specified by its assigned orientation (out of the possible $N_o = 12$ orientations, as specified in Supp. Table~S1 and Supp. Table~S4).  The Boolean variables and clauses are listed in Supp. Table~S1 and described in Section SI in Supplementary Information.

We developed a modified version of the MiniSAT SAT-solver tool \cite{een2005minisat} to enumerate all possible solutions for a given combination of $N_s$ and $N_c$ that can form a DC lattice, and then tests all generated solutions for their ability to assemble a specified undesired lattice (HD 32-unit cell in our case) to quickly filter out the desired solution that can form the DC unit cell but is unable to be arranged into the HD unit cell in a way that satisfies all bonds.

\subsection{DNA System Simulations}
We setup the oxDNA simulations starting from a finite cluster of 128 particles (Fig.~\ref{fig_experimental}), where the unit cell of 16 elements is repeated two times along each axis. The Athena \cite{athena} and oxView \cite{oxView} design tools were used to construct a plausible tetrahedral DNA origami monomer which is aligned to the idealized lattice positions of the cluster using the Biopython SVDSuperimposer package \cite{biopython}. 
Prior to molecular dynamics simulation, the presassembled system needs to be relaxed to correct for possible overlaps or overstretched bonds that were created during the placing of the DNA origamis onto the patchy particle representation. For each of the paired bases in the system an additional pairwise attractive potential was generated using the 
generate\textunderscore force script of the oxDNA\textunderscore analysis\textunderscore tools package \cite{oxView}  with the stiffness parameter of the trap set to stiff = 3. This potential is active during the entire relaxation process.
For simulations we chose the oxDNA2 model with averaged sequence interaction strength and a salt concentration of 1 M. 
The system was first relaxed using a short Monte Carlo simulation in the canonical NVT ensemble with T = 300K.
The relaxation then continued 
using MD simulation for 7\,400\,000 steps with time step $dt = 0.001$
using the Bussi Thermostat \cite{bussi2007canonical} with the $\tau$ value set to 1, T = 293K and oxDNA's options max\textunderscore backbone\textunderscore force = 10, max\textunderscore density\textunderscore multiplier = 4. 
Next, we equilibrated the system without the max\textunderscore backbone\textunderscore force parameter and the pairwise attractive potential for $175\,400\,000$ simulation steps using the Langevin thermostat. The final production run performed $142\,400\,000$ MD steps which, taking into account the value of the diffusion constant set in the simulation ($D_0 = 1$ in oxDNA units), corresponds to about 0.6 ms \cite{snodin2016direct}. 

\subsection{Mean Structure Calculation}
The oxDNA\textunderscore analysis\textunderscore tools~\cite{oxView} scripts were used to obtain a mean structure of the production run: every configuration in the production trajectory is aligned to a randomly chosen configuration from the production run using the Biopython SVDSuperimposer package~\cite{biopython}. This is done to maximize overlap between the sampled states. Next we compute an average of all coordinates describing the system. The final result is in an artificial structure that is a good global descriptor of the analysed trajectory.

\subsection{Patchy Particle Simulations}
In this work, we employ a generalization of the Kern-Frenkel potential \cite{kern2003fluid} to describe the directional interaction between patches. While the Kern-Frenkel potential is discontinous and designed for use in Monte Carlo simulations, we introduce here its differentiable version (dKF potential) that makes it possible to implement it with a Molecular Dynamics (MD) code for simulation of the assembly kinetics, and also allows for efficient implementation on GPUs, thus allowing us to study much longer timescales and larger systems sizes of the nucleating system.

Patchy particles feel a mutual repulsion modelled through a WCA interaction~\cite{weeks::jcp::54::1971}:

\begin{equation}
	U_{ij}(r) =\left\{ 
	\begin{array}{l l}
	4\epsilon \left[ \left( \frac{ \sigma }{r} \right)^{12} - \left( \frac{ \sigma }{r} \right)^{6} + \frac{1}{4}\right] & r \le 2^{\frac{1}{6}} \sigma\\
	0 & r > 2^{\frac{1}{6}} \sigma
	\end{array} \right.
	\label{eq:wca}
\end{equation}

\noindent where $r$ is the centre-to-centre particle distance, $\epsilon$ is the energy scale and $\sigma$ is the particle diameter. We take the latter two quantities as units of measurements of energy and length, respectively.

The patch-patch interaction is a square-well-like attractive potential modulated by an orientation-dependent function. The range of the interaction is set by the parameter $\delta$, while its (half) angular width is controlled by $\cos \theta_{\rm max}$. The interaction between patch $i$ on particle $\alpha$, identified by the unit vector $\hat{\alpha}_i$, and patch $j$ on particle $\beta$, identified by the versor $\hat{\beta}_j$, is given by
\begin{multline}
V_{\rm pp}(\vec{r_{\rm pp}}, \hat{\alpha}_i, \hat{\beta}_j) = \\ -\epsilon \exp\left(-\frac{1}{2} \left( \frac{r_{\rm pp} - \sigma_c}{\delta} \right)^{10}\right) \Omega(-\hat{r}, \hat{\alpha_i}) \Omega(\hat{r}_{\rm pp}, \hat{\beta_j})
\end{multline}
where $\vec{r}_{\rm pp} = \vec{r}_\alpha - \vec{r}_\beta$, $r_{\rm pp} = |\vec{r}_{\rm pp}|$, $\hat{r}_{\rm pp} = \vec{r}_{\rm pp}/r_{\rm pp}$ and $\Omega$ is a steep modulating function that takes into account the orientation of a patch with respect to the unit vector connecting the particles' centres and takes the following form:
\begin{equation}
\Omega(\hat{r}, \hat{\gamma_k}) = \exp\left(-\frac{1}{2} \left( \frac{1 - \hat{r}\cdot\hat{\gamma}_k}{1 - \cos \theta_{\rm max}} \right)^{20}\right).
\end{equation}

In this work we set $\delta = 0.2$ and $\cos \theta_{\rm max} = 0.97$ so that patches can be involved in no more than one bond at a time.

\bibliography{biblio}

\begin{thebibliography}{48}%
\makeatletter
\providecommand \@ifxundefined [1]{%
 \@ifx{#1\undefined}
}%
\providecommand \@ifnum [1]{%
 \ifnum #1\expandafter \@firstoftwo
 \else \expandafter \@secondoftwo
 \fi
}%
\providecommand \@ifx [1]{%
 \ifx #1\expandafter \@firstoftwo
 \else \expandafter \@secondoftwo
 \fi
}%
\providecommand \natexlab [1]{#1}%
\providecommand \enquote  [1]{``#1''}%
\providecommand \bibnamefont  [1]{#1}%
\providecommand \bibfnamefont [1]{#1}%
\providecommand \citenamefont [1]{#1}%
\providecommand \href@noop [0]{\@secondoftwo}%
\providecommand \href [0]{\begingroup \@sanitize@url \@href}%
\providecommand \@href[1]{\@@startlink{#1}\@@href}%
\providecommand \@@href[1]{\endgroup#1\@@endlink}%
\providecommand \@sanitize@url [0]{\catcode `\\12\catcode `\$12\catcode
  `\&12\catcode `\#12\catcode `\^12\catcode `\_12\catcode `\%12\relax}%
\providecommand \@@startlink[1]{}%
\providecommand \@@endlink[0]{}%
\providecommand \url  [0]{\begingroup\@sanitize@url \@url }%
\providecommand \@url [1]{\endgroup\@href {#1}{\urlprefix }}%
\providecommand \urlprefix  [0]{URL }%
\providecommand \Eprint [0]{\href }%
\providecommand \doibase [0]{https://doi.org/}%
\providecommand \selectlanguage [0]{\@gobble}%
\providecommand \bibinfo  [0]{\@secondoftwo}%
\providecommand \bibfield  [0]{\@secondoftwo}%
\providecommand \translation [1]{[#1]}%
\providecommand \BibitemOpen [0]{}%
\providecommand \bibitemStop [0]{}%
\providecommand \bibitemNoStop [0]{.\EOS\space}%
\providecommand \EOS [0]{\spacefactor3000\relax}%
\providecommand \BibitemShut  [1]{\csname bibitem#1\endcsname}%
\let\auto@bib@innerbib\@empty
\bibitem [{\citenamefont {Feynman}(2018)}]{feynman2018there}%
  \BibitemOpen
  \bibfield  {author} {\bibinfo {author} {\bibfnamefont {R.}~\bibnamefont
  {Feynman}},\ }\bibfield  {title} {\bibinfo {title} {There’s plenty of room
  at the bottom},\ }in\ \href@noop {} {\emph {\bibinfo {booktitle} {Feynman and
  computation}}}\ (\bibinfo  {publisher} {CRC Press},\ \bibinfo {year} {2018})\
  pp.\ \bibinfo {pages} {63--76}\BibitemShut {NoStop}%
\bibitem [{\citenamefont {Soukoulis}\ and\ \citenamefont
  {Wegener}(2010)}]{soukoulis2010optical}%
  \BibitemOpen
  \bibfield  {author} {\bibinfo {author} {\bibfnamefont {C.~M.}\ \bibnamefont
  {Soukoulis}}\ and\ \bibinfo {author} {\bibfnamefont {M.}~\bibnamefont
  {Wegener}},\ }\bibfield  {title} {\bibinfo {title} {Optical
  metamaterials—more bulky and less lossy},\ }\href@noop {} {\bibfield
  {journal} {\bibinfo  {journal} {Science}\ }\textbf {\bibinfo {volume}
  {330}},\ \bibinfo {pages} {1633} (\bibinfo {year} {2010})}\BibitemShut
  {NoStop}%
\bibitem [{\citenamefont {Ngo}\ \emph {et~al.}(2006)\citenamefont {Ngo},
  \citenamefont {Liddell}, \citenamefont {Ghebrebrhan},\ and\ \citenamefont
  {Joannopoulos}}]{ngo2006tetrastack}%
  \BibitemOpen
  \bibfield  {author} {\bibinfo {author} {\bibfnamefont {T.}~\bibnamefont
  {Ngo}}, \bibinfo {author} {\bibfnamefont {C.}~\bibnamefont {Liddell}},
  \bibinfo {author} {\bibfnamefont {M.}~\bibnamefont {Ghebrebrhan}},\ and\
  \bibinfo {author} {\bibfnamefont {J.}~\bibnamefont {Joannopoulos}},\
  }\bibfield  {title} {\bibinfo {title} {Tetrastack: Colloidal diamond-inspired
  structure with omnidirectional photonic band gap for low refractive index
  contrast},\ }\href@noop {} {\bibfield  {journal} {\bibinfo  {journal}
  {Applied physics letters}\ }\textbf {\bibinfo {volume} {88}},\ \bibinfo
  {pages} {241920} (\bibinfo {year} {2006})}\BibitemShut {NoStop}%
\bibitem [{\citenamefont {Werts}\ \emph {et~al.}(2002)\citenamefont {Werts},
  \citenamefont {Lambert}, \citenamefont {Bourgoin},\ and\ \citenamefont
  {Brust}}]{werts2002nanometer}%
  \BibitemOpen
  \bibfield  {author} {\bibinfo {author} {\bibfnamefont {M.~H.}\ \bibnamefont
  {Werts}}, \bibinfo {author} {\bibfnamefont {M.}~\bibnamefont {Lambert}},
  \bibinfo {author} {\bibfnamefont {J.-P.}\ \bibnamefont {Bourgoin}},\ and\
  \bibinfo {author} {\bibfnamefont {M.}~\bibnamefont {Brust}},\ }\bibfield
  {title} {\bibinfo {title} {Nanometer scale patterning of langmuir- blodgett
  films of gold nanoparticles by electron beam lithography},\ }\href@noop {}
  {\bibfield  {journal} {\bibinfo  {journal} {Nano Letters}\ }\textbf {\bibinfo
  {volume} {2}},\ \bibinfo {pages} {43} (\bibinfo {year} {2002})}\BibitemShut
  {NoStop}%
\bibitem [{\citenamefont {Ducrot}\ \emph {et~al.}(2017)\citenamefont {Ducrot},
  \citenamefont {He}, \citenamefont {Yi},\ and\ \citenamefont
  {Pine}}]{ducrot2017colloidal}%
  \BibitemOpen
  \bibfield  {author} {\bibinfo {author} {\bibfnamefont {{\'E}.}~\bibnamefont
  {Ducrot}}, \bibinfo {author} {\bibfnamefont {M.}~\bibnamefont {He}}, \bibinfo
  {author} {\bibfnamefont {G.-R.}\ \bibnamefont {Yi}},\ and\ \bibinfo {author}
  {\bibfnamefont {D.~J.}\ \bibnamefont {Pine}},\ }\bibfield  {title} {\bibinfo
  {title} {Colloidal alloys with preassembled clusters and spheres},\
  }\href@noop {} {\bibfield  {journal} {\bibinfo  {journal} {Nature materials}\
  }\textbf {\bibinfo {volume} {16}},\ \bibinfo {pages} {652} (\bibinfo {year}
  {2017})}\BibitemShut {NoStop}%
\bibitem [{\citenamefont {Liu}\ \emph {et~al.}(2016)\citenamefont {Liu},
  \citenamefont {Tagawa}, \citenamefont {Xin}, \citenamefont {Wang},
  \citenamefont {Emamy}, \citenamefont {Li}, \citenamefont {Yager},
  \citenamefont {Starr}, \citenamefont {Tkachenko},\ and\ \citenamefont
  {Gang}}]{liu2016diamond}%
  \BibitemOpen
  \bibfield  {author} {\bibinfo {author} {\bibfnamefont {W.}~\bibnamefont
  {Liu}}, \bibinfo {author} {\bibfnamefont {M.}~\bibnamefont {Tagawa}},
  \bibinfo {author} {\bibfnamefont {H.~L.}\ \bibnamefont {Xin}}, \bibinfo
  {author} {\bibfnamefont {T.}~\bibnamefont {Wang}}, \bibinfo {author}
  {\bibfnamefont {H.}~\bibnamefont {Emamy}}, \bibinfo {author} {\bibfnamefont
  {H.}~\bibnamefont {Li}}, \bibinfo {author} {\bibfnamefont {K.~G.}\
  \bibnamefont {Yager}}, \bibinfo {author} {\bibfnamefont {F.~W.}\ \bibnamefont
  {Starr}}, \bibinfo {author} {\bibfnamefont {A.~V.}\ \bibnamefont
  {Tkachenko}},\ and\ \bibinfo {author} {\bibfnamefont {O.}~\bibnamefont
  {Gang}},\ }\bibfield  {title} {\bibinfo {title} {Diamond family of
  nanoparticle superlattices},\ }\href@noop {} {\bibfield  {journal} {\bibinfo
  {journal} {Science}\ }\textbf {\bibinfo {volume} {351}},\ \bibinfo {pages}
  {582} (\bibinfo {year} {2016})}\BibitemShut {NoStop}%
\bibitem [{\citenamefont {Michelson}\ \emph {et~al.}(2022)\citenamefont
  {Michelson}, \citenamefont {Minevich}, \citenamefont {Emamy}, \citenamefont
  {Huang}, \citenamefont {Chu}, \citenamefont {Yan},\ and\ \citenamefont
  {Gang}}]{michelson2022three}%
  \BibitemOpen
  \bibfield  {author} {\bibinfo {author} {\bibfnamefont {A.}~\bibnamefont
  {Michelson}}, \bibinfo {author} {\bibfnamefont {B.}~\bibnamefont {Minevich}},
  \bibinfo {author} {\bibfnamefont {H.}~\bibnamefont {Emamy}}, \bibinfo
  {author} {\bibfnamefont {X.}~\bibnamefont {Huang}}, \bibinfo {author}
  {\bibfnamefont {Y.~S.}\ \bibnamefont {Chu}}, \bibinfo {author} {\bibfnamefont
  {H.}~\bibnamefont {Yan}},\ and\ \bibinfo {author} {\bibfnamefont
  {O.}~\bibnamefont {Gang}},\ }\bibfield  {title} {\bibinfo {title}
  {Three-dimensional visualization of nanoparticle lattices and multimaterial
  frameworks},\ }\href@noop {} {\bibfield  {journal} {\bibinfo  {journal}
  {Science}\ }\textbf {\bibinfo {volume} {376}},\ \bibinfo {pages} {203}
  (\bibinfo {year} {2022})}\BibitemShut {NoStop}%
\bibitem [{\citenamefont {He}\ \emph {et~al.}(2020)\citenamefont {He},
  \citenamefont {Gales}, \citenamefont {Ducrot}, \citenamefont {Gong},
  \citenamefont {Yi}, \citenamefont {Sacanna},\ and\ \citenamefont
  {Pine}}]{he2020colloidal}%
  \BibitemOpen
  \bibfield  {author} {\bibinfo {author} {\bibfnamefont {M.}~\bibnamefont
  {He}}, \bibinfo {author} {\bibfnamefont {J.~P.}\ \bibnamefont {Gales}},
  \bibinfo {author} {\bibfnamefont {{\'E}.}~\bibnamefont {Ducrot}}, \bibinfo
  {author} {\bibfnamefont {Z.}~\bibnamefont {Gong}}, \bibinfo {author}
  {\bibfnamefont {G.-R.}\ \bibnamefont {Yi}}, \bibinfo {author} {\bibfnamefont
  {S.}~\bibnamefont {Sacanna}},\ and\ \bibinfo {author} {\bibfnamefont {D.~J.}\
  \bibnamefont {Pine}},\ }\bibfield  {title} {\bibinfo {title} {Colloidal
  diamond},\ }\href@noop {} {\bibfield  {journal} {\bibinfo  {journal}
  {Nature}\ }\textbf {\bibinfo {volume} {585}},\ \bibinfo {pages} {524}
  (\bibinfo {year} {2020})}\BibitemShut {NoStop}%
\bibitem [{\citenamefont {Russo}\ \emph {et~al.}(2021)\citenamefont {Russo},
  \citenamefont {Leoni}, \citenamefont {Martelli},\ and\ \citenamefont
  {Sciortino}}]{russo2021physics}%
  \BibitemOpen
  \bibfield  {author} {\bibinfo {author} {\bibfnamefont {J.}~\bibnamefont
  {Russo}}, \bibinfo {author} {\bibfnamefont {F.}~\bibnamefont {Leoni}},
  \bibinfo {author} {\bibfnamefont {F.}~\bibnamefont {Martelli}},\ and\
  \bibinfo {author} {\bibfnamefont {F.}~\bibnamefont {Sciortino}},\ }\bibfield
  {title} {\bibinfo {title} {The physics of empty liquids: from patchy
  particles to water},\ }\href@noop {} {\bibfield  {journal} {\bibinfo
  {journal} {Reports on Progress in Physics}\ } (\bibinfo {year}
  {2021})}\BibitemShut {NoStop}%
\bibitem [{\citenamefont {van Anders}\ \emph {et~al.}(2013)\citenamefont {van
  Anders}, \citenamefont {Ahmed}, \citenamefont {Smith}, \citenamefont
  {Engel},\ and\ \citenamefont {Glotzer}}]{van2013entropically}%
  \BibitemOpen
  \bibfield  {author} {\bibinfo {author} {\bibfnamefont {G.}~\bibnamefont {van
  Anders}}, \bibinfo {author} {\bibfnamefont {N.~K.}\ \bibnamefont {Ahmed}},
  \bibinfo {author} {\bibfnamefont {R.}~\bibnamefont {Smith}}, \bibinfo
  {author} {\bibfnamefont {M.}~\bibnamefont {Engel}},\ and\ \bibinfo {author}
  {\bibfnamefont {S.~C.}\ \bibnamefont {Glotzer}},\ }\bibfield  {title}
  {\bibinfo {title} {Entropically patchy particles: engineering valence through
  shape entropy},\ }\href@noop {} {\bibfield  {journal} {\bibinfo  {journal}
  {Acs Nano}\ }\textbf {\bibinfo {volume} {8}},\ \bibinfo {pages} {931}
  (\bibinfo {year} {2013})}\BibitemShut {NoStop}%
\bibitem [{\citenamefont {Zhang}\ and\ \citenamefont
  {Glotzer}(2004)}]{zhang2004self}%
  \BibitemOpen
  \bibfield  {author} {\bibinfo {author} {\bibfnamefont {Z.}~\bibnamefont
  {Zhang}}\ and\ \bibinfo {author} {\bibfnamefont {S.~C.}\ \bibnamefont
  {Glotzer}},\ }\bibfield  {title} {\bibinfo {title} {Self-assembly of patchy
  particles},\ }\href@noop {} {\bibfield  {journal} {\bibinfo  {journal} {Nano
  Letters}\ }\textbf {\bibinfo {volume} {4}},\ \bibinfo {pages} {1407}
  (\bibinfo {year} {2004})}\BibitemShut {NoStop}%
\bibitem [{\citenamefont {Pawar}\ and\ \citenamefont
  {Kretzschmar}(2010)}]{pawar2010fabrication}%
  \BibitemOpen
  \bibfield  {author} {\bibinfo {author} {\bibfnamefont {A.~B.}\ \bibnamefont
  {Pawar}}\ and\ \bibinfo {author} {\bibfnamefont {I.}~\bibnamefont
  {Kretzschmar}},\ }\bibfield  {title} {\bibinfo {title} {Fabrication,
  assembly, and application of patchy particles},\ }\href@noop {} {\bibfield
  {journal} {\bibinfo  {journal} {Macromolecular rapid communications}\
  }\textbf {\bibinfo {volume} {31}},\ \bibinfo {pages} {150} (\bibinfo {year}
  {2010})}\BibitemShut {NoStop}%
\bibitem [{\citenamefont {Bianchi}\ \emph {et~al.}(2011)\citenamefont
  {Bianchi}, \citenamefont {Blaak},\ and\ \citenamefont
  {Likos}}]{bianchi2011patchy}%
  \BibitemOpen
  \bibfield  {author} {\bibinfo {author} {\bibfnamefont {E.}~\bibnamefont
  {Bianchi}}, \bibinfo {author} {\bibfnamefont {R.}~\bibnamefont {Blaak}},\
  and\ \bibinfo {author} {\bibfnamefont {C.~N.}\ \bibnamefont {Likos}},\
  }\bibfield  {title} {\bibinfo {title} {Patchy colloids: state of the art and
  perspectives},\ }\href@noop {} {\bibfield  {journal} {\bibinfo  {journal}
  {Physical Chemistry Chemical Physics}\ }\textbf {\bibinfo {volume} {13}},\
  \bibinfo {pages} {6397} (\bibinfo {year} {2011})}\BibitemShut {NoStop}%
\bibitem [{\citenamefont {Romano}\ and\ \citenamefont
  {Sciortino}(2011)}]{romano2011colloidal}%
  \BibitemOpen
  \bibfield  {author} {\bibinfo {author} {\bibfnamefont {F.}~\bibnamefont
  {Romano}}\ and\ \bibinfo {author} {\bibfnamefont {F.}~\bibnamefont
  {Sciortino}},\ }\bibfield  {title} {\bibinfo {title} {Colloidal
  self-assembly: patchy from the bottom up},\ }\href@noop {} {\bibfield
  {journal} {\bibinfo  {journal} {Nature materials}\ }\textbf {\bibinfo
  {volume} {10}},\ \bibinfo {pages} {171} (\bibinfo {year} {2011})}\BibitemShut
  {NoStop}%
\bibitem [{\citenamefont {Suzuki}\ \emph {et~al.}(2009)\citenamefont {Suzuki},
  \citenamefont {Hosokawa},\ and\ \citenamefont
  {Maeda}}]{suzuki2009controlling}%
  \BibitemOpen
  \bibfield  {author} {\bibinfo {author} {\bibfnamefont {K.}~\bibnamefont
  {Suzuki}}, \bibinfo {author} {\bibfnamefont {K.}~\bibnamefont {Hosokawa}},\
  and\ \bibinfo {author} {\bibfnamefont {M.}~\bibnamefont {Maeda}},\ }\bibfield
   {title} {\bibinfo {title} {Controlling the number and positions of
  oligonucleotides on gold nanoparticle surfaces},\ }\href@noop {} {\bibfield
  {journal} {\bibinfo  {journal} {Journal of the American Chemical Society}\
  }\textbf {\bibinfo {volume} {131}},\ \bibinfo {pages} {7518} (\bibinfo {year}
  {2009})}\BibitemShut {NoStop}%
\bibitem [{\citenamefont {Kim}\ \emph {et~al.}(2011)\citenamefont {Kim},
  \citenamefont {Kim},\ and\ \citenamefont {Deaton}}]{kim2011dna}%
  \BibitemOpen
  \bibfield  {author} {\bibinfo {author} {\bibfnamefont {J.-W.}\ \bibnamefont
  {Kim}}, \bibinfo {author} {\bibfnamefont {J.-H.}\ \bibnamefont {Kim}},\ and\
  \bibinfo {author} {\bibfnamefont {R.}~\bibnamefont {Deaton}},\ }\bibfield
  {title} {\bibinfo {title} {Dna-linked nanoparticle building blocks for
  programmable matter},\ }\href@noop {} {\bibfield  {journal} {\bibinfo
  {journal} {Angewandte Chemie International Edition}\ }\textbf {\bibinfo
  {volume} {50}},\ \bibinfo {pages} {9185} (\bibinfo {year}
  {2011})}\BibitemShut {NoStop}%
\bibitem [{\citenamefont {Wang}\ \emph {et~al.}(2012)\citenamefont {Wang},
  \citenamefont {Wang}, \citenamefont {Breed}, \citenamefont {Manoharan},
  \citenamefont {Feng}, \citenamefont {Hollingsworth}, \citenamefont {Weck},\
  and\ \citenamefont {Pine}}]{wang2012colloids}%
  \BibitemOpen
  \bibfield  {author} {\bibinfo {author} {\bibfnamefont {Y.}~\bibnamefont
  {Wang}}, \bibinfo {author} {\bibfnamefont {Y.}~\bibnamefont {Wang}}, \bibinfo
  {author} {\bibfnamefont {D.~R.}\ \bibnamefont {Breed}}, \bibinfo {author}
  {\bibfnamefont {V.~N.}\ \bibnamefont {Manoharan}}, \bibinfo {author}
  {\bibfnamefont {L.}~\bibnamefont {Feng}}, \bibinfo {author} {\bibfnamefont
  {A.~D.}\ \bibnamefont {Hollingsworth}}, \bibinfo {author} {\bibfnamefont
  {M.}~\bibnamefont {Weck}},\ and\ \bibinfo {author} {\bibfnamefont {D.~J.}\
  \bibnamefont {Pine}},\ }\bibfield  {title} {\bibinfo {title} {Colloids with
  valence and specific directional bonding},\ }\href@noop {} {\bibfield
  {journal} {\bibinfo  {journal} {Nature}\ }\textbf {\bibinfo {volume} {491}},\
  \bibinfo {pages} {51} (\bibinfo {year} {2012})}\BibitemShut {NoStop}%
\bibitem [{\citenamefont {Feng}\ \emph {et~al.}(2013)\citenamefont {Feng},
  \citenamefont {Dreyfus}, \citenamefont {Sha}, \citenamefont {Seeman},\ and\
  \citenamefont {Chaikin}}]{feng2013dna}%
  \BibitemOpen
  \bibfield  {author} {\bibinfo {author} {\bibfnamefont {L.}~\bibnamefont
  {Feng}}, \bibinfo {author} {\bibfnamefont {R.}~\bibnamefont {Dreyfus}},
  \bibinfo {author} {\bibfnamefont {R.}~\bibnamefont {Sha}}, \bibinfo {author}
  {\bibfnamefont {N.~C.}\ \bibnamefont {Seeman}},\ and\ \bibinfo {author}
  {\bibfnamefont {P.~M.}\ \bibnamefont {Chaikin}},\ }\bibfield  {title}
  {\bibinfo {title} {Dna patchy particles},\ }\href@noop {} {\bibfield
  {journal} {\bibinfo  {journal} {Advanced Materials}\ }\textbf {\bibinfo
  {volume} {25}},\ \bibinfo {pages} {2779} (\bibinfo {year}
  {2013})}\BibitemShut {NoStop}%
\bibitem [{\citenamefont {Mahynski}\ \emph {et~al.}(2016)\citenamefont
  {Mahynski}, \citenamefont {Rovigatti}, \citenamefont {Likos},\ and\
  \citenamefont {Panagiotopoulos}}]{mahynski2016bottom}%
  \BibitemOpen
  \bibfield  {author} {\bibinfo {author} {\bibfnamefont {N.~A.}\ \bibnamefont
  {Mahynski}}, \bibinfo {author} {\bibfnamefont {L.}~\bibnamefont {Rovigatti}},
  \bibinfo {author} {\bibfnamefont {C.~N.}\ \bibnamefont {Likos}},\ and\
  \bibinfo {author} {\bibfnamefont {A.~Z.}\ \bibnamefont {Panagiotopoulos}},\
  }\bibfield  {title} {\bibinfo {title} {Bottom-up colloidal crystal assembly
  with a twist},\ }\href@noop {} {\bibfield  {journal} {\bibinfo  {journal}
  {ACS nano}\ }\textbf {\bibinfo {volume} {10}},\ \bibinfo {pages} {5459}
  (\bibinfo {year} {2016})}\BibitemShut {NoStop}%
\bibitem [{\citenamefont {Leoni}\ and\ \citenamefont
  {Russo}(2021)}]{leoni2021non}%
  \BibitemOpen
  \bibfield  {author} {\bibinfo {author} {\bibfnamefont {F.}~\bibnamefont
  {Leoni}}\ and\ \bibinfo {author} {\bibfnamefont {J.}~\bibnamefont {Russo}},\
  }\bibfield  {title} {\bibinfo {title} {Non-classical nucleation pathways in
  stacking-disordered crystals},\ }\href@noop {} {\bibfield  {journal}
  {\bibinfo  {journal} {arXiv preprint arXiv:2105.05506}\ } (\bibinfo {year}
  {2021})}\BibitemShut {NoStop}%
\bibitem [{\citenamefont {Romano}\ and\ \citenamefont
  {Sciortino}(2012)}]{romano2012patterning}%
  \BibitemOpen
  \bibfield  {author} {\bibinfo {author} {\bibfnamefont {F.}~\bibnamefont
  {Romano}}\ and\ \bibinfo {author} {\bibfnamefont {F.}~\bibnamefont
  {Sciortino}},\ }\bibfield  {title} {\bibinfo {title} {Patterning symmetry in
  the rational design of colloidal crystals},\ }\href@noop {} {\bibfield
  {journal} {\bibinfo  {journal} {Nature communications}\ }\textbf {\bibinfo
  {volume} {3}},\ \bibinfo {pages} {975} (\bibinfo {year} {2012})}\BibitemShut
  {NoStop}%
\bibitem [{\citenamefont {Tracey}\ \emph {et~al.}(2019)\citenamefont {Tracey},
  \citenamefont {Noya},\ and\ \citenamefont {Doye}}]{tracey2019programming}%
  \BibitemOpen
  \bibfield  {author} {\bibinfo {author} {\bibfnamefont {D.~F.}\ \bibnamefont
  {Tracey}}, \bibinfo {author} {\bibfnamefont {E.~G.}\ \bibnamefont {Noya}},\
  and\ \bibinfo {author} {\bibfnamefont {J.~P.~K.}\ \bibnamefont {Doye}},\
  }\bibfield  {title} {\bibinfo {title} {Programming patchy particles to form
  complex periodic structures},\ }\href@noop {} {\bibfield  {journal} {\bibinfo
   {journal} {The Journal of Chemical Physics}\ }\textbf {\bibinfo {volume}
  {151}},\ \bibinfo {pages} {224506} (\bibinfo {year} {2019})}\BibitemShut
  {NoStop}%
\bibitem [{\citenamefont {Morphew}\ \emph {et~al.}(2018)\citenamefont
  {Morphew}, \citenamefont {Shaw}, \citenamefont {Avins},\ and\ \citenamefont
  {Chakrabarti}}]{morphew2018programming}%
  \BibitemOpen
  \bibfield  {author} {\bibinfo {author} {\bibfnamefont {D.}~\bibnamefont
  {Morphew}}, \bibinfo {author} {\bibfnamefont {J.}~\bibnamefont {Shaw}},
  \bibinfo {author} {\bibfnamefont {C.}~\bibnamefont {Avins}},\ and\ \bibinfo
  {author} {\bibfnamefont {D.}~\bibnamefont {Chakrabarti}},\ }\bibfield
  {title} {\bibinfo {title} {Programming hierarchical self-assembly of patchy
  particles into colloidal crystals via colloidal molecules},\ }\href@noop {}
  {\bibfield  {journal} {\bibinfo  {journal} {ACS nano}\ }\textbf {\bibinfo
  {volume} {12}},\ \bibinfo {pages} {2355} (\bibinfo {year}
  {2018})}\BibitemShut {NoStop}%
\bibitem [{\citenamefont {Patra}\ and\ \citenamefont
  {Tkachenko}(2018)}]{patra2018programmable}%
  \BibitemOpen
  \bibfield  {author} {\bibinfo {author} {\bibfnamefont {N.}~\bibnamefont
  {Patra}}\ and\ \bibinfo {author} {\bibfnamefont {A.~V.}\ \bibnamefont
  {Tkachenko}},\ }\bibfield  {title} {\bibinfo {title} {Programmable
  self-assembly of diamond polymorphs from chromatic patchy particles},\
  }\href@noop {} {\bibfield  {journal} {\bibinfo  {journal} {Physical Review
  E}\ }\textbf {\bibinfo {volume} {98}},\ \bibinfo {pages} {032611} (\bibinfo
  {year} {2018})}\BibitemShut {NoStop}%
\bibitem [{\citenamefont {Ma}\ and\ \citenamefont
  {Ferguson}(2019)}]{ma2019inverse}%
  \BibitemOpen
  \bibfield  {author} {\bibinfo {author} {\bibfnamefont {Y.}~\bibnamefont
  {Ma}}\ and\ \bibinfo {author} {\bibfnamefont {A.~L.}\ \bibnamefont
  {Ferguson}},\ }\bibfield  {title} {\bibinfo {title} {Inverse design of
  self-assembling colloidal crystals with omnidirectional photonic bandgaps},\
  }\href@noop {} {\bibfield  {journal} {\bibinfo  {journal} {Soft matter}\
  }\textbf {\bibinfo {volume} {15}},\ \bibinfo {pages} {8808} (\bibinfo {year}
  {2019})}\BibitemShut {NoStop}%
\bibitem [{\citenamefont {Neophytou}\ \emph {et~al.}(2021)\citenamefont
  {Neophytou}, \citenamefont {Chakrabarti},\ and\ \citenamefont
  {Sciortino}}]{neophytou2021facile}%
  \BibitemOpen
  \bibfield  {author} {\bibinfo {author} {\bibfnamefont {A.}~\bibnamefont
  {Neophytou}}, \bibinfo {author} {\bibfnamefont {D.}~\bibnamefont
  {Chakrabarti}},\ and\ \bibinfo {author} {\bibfnamefont {F.}~\bibnamefont
  {Sciortino}},\ }\bibfield  {title} {\bibinfo {title} {Facile self-assembly of
  colloidal diamond from tetrahedral patchy particles via ring selection},\
  }\href@noop {} {\bibfield  {journal} {\bibinfo  {journal} {Proceedings of the
  National Academy of Sciences}\ }\textbf {\bibinfo {volume} {118}} (\bibinfo
  {year} {2021})}\BibitemShut {NoStop}%
\bibitem [{\citenamefont {Bisker}\ and\ \citenamefont
  {England}(2018)}]{bisker2018nonequilibrium}%
  \BibitemOpen
  \bibfield  {author} {\bibinfo {author} {\bibfnamefont {G.}~\bibnamefont
  {Bisker}}\ and\ \bibinfo {author} {\bibfnamefont {J.~L.}\ \bibnamefont
  {England}},\ }\bibfield  {title} {\bibinfo {title} {Nonequilibrium
  associative retrieval of multiple stored self-assembly targets},\ }\href@noop
  {} {\bibfield  {journal} {\bibinfo  {journal} {Proceedings of the National
  Academy of Sciences}\ }\textbf {\bibinfo {volume} {115}},\ \bibinfo {pages}
  {E10531} (\bibinfo {year} {2018})}\BibitemShut {NoStop}%
\bibitem [{\citenamefont {Whitelam}\ and\ \citenamefont
  {Tamblyn}(2020)}]{whitelam2020learning}%
  \BibitemOpen
  \bibfield  {author} {\bibinfo {author} {\bibfnamefont {S.}~\bibnamefont
  {Whitelam}}\ and\ \bibinfo {author} {\bibfnamefont {I.}~\bibnamefont
  {Tamblyn}},\ }\bibfield  {title} {\bibinfo {title} {Learning to grow: Control
  of material self-assembly using evolutionary reinforcement learning},\
  }\href@noop {} {\bibfield  {journal} {\bibinfo  {journal} {Physical Review
  E}\ }\textbf {\bibinfo {volume} {101}},\ \bibinfo {pages} {052604} (\bibinfo
  {year} {2020})}\BibitemShut {NoStop}%
\bibitem [{\citenamefont {Whitelam}\ and\ \citenamefont
  {Tamblyn}(2021)}]{whitelam2021neuroevolutionary}%
  \BibitemOpen
  \bibfield  {author} {\bibinfo {author} {\bibfnamefont {S.}~\bibnamefont
  {Whitelam}}\ and\ \bibinfo {author} {\bibfnamefont {I.}~\bibnamefont
  {Tamblyn}},\ }\bibfield  {title} {\bibinfo {title} {Neuroevolutionary
  learning of particles and protocols for self-assembly},\ }\href@noop {}
  {\bibfield  {journal} {\bibinfo  {journal} {Physical review letters}\
  }\textbf {\bibinfo {volume} {127}},\ \bibinfo {pages} {018003} (\bibinfo
  {year} {2021})}\BibitemShut {NoStop}%
\bibitem [{\citenamefont {Bupathy}\ \emph {et~al.}(2021)\citenamefont
  {Bupathy}, \citenamefont {Frenkel},\ and\ \citenamefont
  {Sastry}}]{bupathy2021temperature}%
  \BibitemOpen
  \bibfield  {author} {\bibinfo {author} {\bibfnamefont {A.}~\bibnamefont
  {Bupathy}}, \bibinfo {author} {\bibfnamefont {D.}~\bibnamefont {Frenkel}},\
  and\ \bibinfo {author} {\bibfnamefont {S.}~\bibnamefont {Sastry}},\
  }\bibfield  {title} {\bibinfo {title} {Temperature protocols to guide
  selective self-assembly of competing structures},\ }\href@noop {} {\bibfield
  {journal} {\bibinfo  {journal} {arXiv preprint arXiv:2110.11274}\ } (\bibinfo
  {year} {2021})}\BibitemShut {NoStop}%
\bibitem [{\citenamefont {Romano}\ \emph {et~al.}(2020)\citenamefont {Romano},
  \citenamefont {Russo}, \citenamefont {Kroc},\ and\ \citenamefont
  {{\v{S}}ulc}}]{romano2020designing}%
  \BibitemOpen
  \bibfield  {author} {\bibinfo {author} {\bibfnamefont {F.}~\bibnamefont
  {Romano}}, \bibinfo {author} {\bibfnamefont {J.}~\bibnamefont {Russo}},
  \bibinfo {author} {\bibfnamefont {L.}~\bibnamefont {Kroc}},\ and\ \bibinfo
  {author} {\bibfnamefont {P.}~\bibnamefont {{\v{S}}ulc}},\ }\bibfield  {title}
  {\bibinfo {title} {Designing patchy interactions to self-assemble arbitrary
  structures},\ }\href@noop {} {\bibfield  {journal} {\bibinfo  {journal}
  {Physical Review Letters}\ }\textbf {\bibinfo {volume} {125}},\ \bibinfo
  {pages} {118003} (\bibinfo {year} {2020})}\BibitemShut {NoStop}%
\bibitem [{\citenamefont {Russo}\ \emph {et~al.}(2022)\citenamefont {Russo},
  \citenamefont {Romano}, \citenamefont {Kroc}, \citenamefont {Sciortino},
  \citenamefont {Rovigatti},\ and\ \citenamefont {{\v{S}}ulc}}]{russo2021sat}%
  \BibitemOpen
  \bibfield  {author} {\bibinfo {author} {\bibfnamefont {J.}~\bibnamefont
  {Russo}}, \bibinfo {author} {\bibfnamefont {F.}~\bibnamefont {Romano}},
  \bibinfo {author} {\bibfnamefont {L.}~\bibnamefont {Kroc}}, \bibinfo {author}
  {\bibfnamefont {F.}~\bibnamefont {Sciortino}}, \bibinfo {author}
  {\bibfnamefont {L.}~\bibnamefont {Rovigatti}},\ and\ \bibinfo {author}
  {\bibfnamefont {P.}~\bibnamefont {{\v{S}}ulc}},\ }\bibfield  {title}
  {\bibinfo {title} {{SAT}-assembly: A new approach for designing
  self-assembling systems},\ }\href@noop {} {\bibfield  {journal} {\bibinfo
  {journal} {Journal of Physics: Condensed Matter}\ } (\bibinfo {year}
  {2022})}\BibitemShut {NoStop}%
\bibitem [{\citenamefont {Snodin}\ \emph {et~al.}(2016)\citenamefont {Snodin},
  \citenamefont {Romano}, \citenamefont {Rovigatti}, \citenamefont {Ouldridge},
  \citenamefont {Louis},\ and\ \citenamefont {Doye}}]{snodin2016direct}%
  \BibitemOpen
  \bibfield  {author} {\bibinfo {author} {\bibfnamefont {B.~E.}\ \bibnamefont
  {Snodin}}, \bibinfo {author} {\bibfnamefont {F.}~\bibnamefont {Romano}},
  \bibinfo {author} {\bibfnamefont {L.}~\bibnamefont {Rovigatti}}, \bibinfo
  {author} {\bibfnamefont {T.~E.}\ \bibnamefont {Ouldridge}}, \bibinfo {author}
  {\bibfnamefont {A.~A.}\ \bibnamefont {Louis}},\ and\ \bibinfo {author}
  {\bibfnamefont {J.~P.}\ \bibnamefont {Doye}},\ }\bibfield  {title} {\bibinfo
  {title} {Direct simulation of the self-assembly of a small {DNA} origami},\
  }\href@noop {} {\bibfield  {journal} {\bibinfo  {journal} {ACS nano}\
  }\textbf {\bibinfo {volume} {10}},\ \bibinfo {pages} {1724} (\bibinfo {year}
  {2016})}\BibitemShut {NoStop}%
\bibitem [{\citenamefont {Rovigatti}\ \emph {et~al.}(2015)\citenamefont
  {Rovigatti}, \citenamefont {{\v{S}}ulc}, \citenamefont {Reguly},\ and\
  \citenamefont {Romano}}]{rovigatti2015comparison}%
  \BibitemOpen
  \bibfield  {author} {\bibinfo {author} {\bibfnamefont {L.}~\bibnamefont
  {Rovigatti}}, \bibinfo {author} {\bibfnamefont {P.}~\bibnamefont
  {{\v{S}}ulc}}, \bibinfo {author} {\bibfnamefont {I.~Z.}\ \bibnamefont
  {Reguly}},\ and\ \bibinfo {author} {\bibfnamefont {F.}~\bibnamefont
  {Romano}},\ }\bibfield  {title} {\bibinfo {title} {A comparison between
  parallelization approaches in molecular dynamics simulations on gpus},\
  }\href@noop {} {\bibfield  {journal} {\bibinfo  {journal} {Journal of
  computational chemistry}\ }\textbf {\bibinfo {volume} {36}},\ \bibinfo
  {pages} {1} (\bibinfo {year} {2015})}\BibitemShut {NoStop}%
\bibitem [{\citenamefont {Ouldridge}\ \emph {et~al.}(2011)\citenamefont
  {Ouldridge}, \citenamefont {Louis},\ and\ \citenamefont
  {Doye}}]{ouldridge2011structural}%
  \BibitemOpen
  \bibfield  {author} {\bibinfo {author} {\bibfnamefont {T.~E.}\ \bibnamefont
  {Ouldridge}}, \bibinfo {author} {\bibfnamefont {A.~A.}\ \bibnamefont
  {Louis}},\ and\ \bibinfo {author} {\bibfnamefont {J.~P.}\ \bibnamefont
  {Doye}},\ }\bibfield  {title} {\bibinfo {title} {Structural, mechanical, and
  thermodynamic properties of a coarse-grained {DNA} model},\ }\href@noop {}
  {\bibfield  {journal} {\bibinfo  {journal} {The Journal of chemical physics}\
  }\textbf {\bibinfo {volume} {134}},\ \bibinfo {pages} {02B627} (\bibinfo
  {year} {2011})}\BibitemShut {NoStop}%
\bibitem [{\citenamefont {\v{S}ulc}\ \emph {et~al.}(2012)\citenamefont
  {\v{S}ulc}, \citenamefont {Romano}, \citenamefont {Ouldridge}, \citenamefont
  {Rovigatti}, \citenamefont {Doye},\ and\ \citenamefont
  {Louis}}]{sulc2012sequence}%
  \BibitemOpen
  \bibfield  {author} {\bibinfo {author} {\bibfnamefont {P.}~\bibnamefont
  {\v{S}ulc}}, \bibinfo {author} {\bibfnamefont {F.}~\bibnamefont {Romano}},
  \bibinfo {author} {\bibfnamefont {T.~E.}\ \bibnamefont {Ouldridge}}, \bibinfo
  {author} {\bibfnamefont {L.}~\bibnamefont {Rovigatti}}, \bibinfo {author}
  {\bibfnamefont {J.~P.~K.}\ \bibnamefont {Doye}},\ and\ \bibinfo {author}
  {\bibfnamefont {A.~A.}\ \bibnamefont {Louis}},\ }\bibfield  {title} {\bibinfo
  {title} {Sequence-dependent thermodynamics of a coarse-grained {DNA} model},\
  }\href@noop {} {\bibfield  {journal} {\bibinfo  {journal} {Journal of
  Chemical Physics}\ }\textbf {\bibinfo {volume} {137}},\ \bibinfo {pages}
  {5101} (\bibinfo {year} {2012})}\BibitemShut {NoStop}%
\bibitem [{\citenamefont {Xiong}\ \emph {et~al.}(2020)\citenamefont {Xiong},
  \citenamefont {Yang}, \citenamefont {Tian}, \citenamefont {Michelson},
  \citenamefont {Xiang}, \citenamefont {Xin},\ and\ \citenamefont
  {Gang}}]{xiong2020three}%
  \BibitemOpen
  \bibfield  {author} {\bibinfo {author} {\bibfnamefont {Y.}~\bibnamefont
  {Xiong}}, \bibinfo {author} {\bibfnamefont {S.}~\bibnamefont {Yang}},
  \bibinfo {author} {\bibfnamefont {Y.}~\bibnamefont {Tian}}, \bibinfo {author}
  {\bibfnamefont {A.}~\bibnamefont {Michelson}}, \bibinfo {author}
  {\bibfnamefont {S.}~\bibnamefont {Xiang}}, \bibinfo {author} {\bibfnamefont
  {H.}~\bibnamefont {Xin}},\ and\ \bibinfo {author} {\bibfnamefont
  {O.}~\bibnamefont {Gang}},\ }\bibfield  {title} {\bibinfo {title}
  {Three-dimensional patterning of nanoparticles by molecular stamping},\
  }\href@noop {} {\bibfield  {journal} {\bibinfo  {journal} {ACS nano}\
  }\textbf {\bibinfo {volume} {14}},\ \bibinfo {pages} {6823} (\bibinfo {year}
  {2020})}\BibitemShut {NoStop}%
\bibitem [{\citenamefont {Tian}\ \emph {et~al.}(2020)\citenamefont {Tian},
  \citenamefont {Lhermitte}, \citenamefont {Bai}, \citenamefont {Vo},
  \citenamefont {Xin}, \citenamefont {Li}, \citenamefont {Li}, \citenamefont
  {Fukuto}, \citenamefont {Yager}, \citenamefont {Kahn} \emph
  {et~al.}}]{tian2020ordered}%
  \BibitemOpen
  \bibfield  {author} {\bibinfo {author} {\bibfnamefont {Y.}~\bibnamefont
  {Tian}}, \bibinfo {author} {\bibfnamefont {J.~R.}\ \bibnamefont {Lhermitte}},
  \bibinfo {author} {\bibfnamefont {L.}~\bibnamefont {Bai}}, \bibinfo {author}
  {\bibfnamefont {T.}~\bibnamefont {Vo}}, \bibinfo {author} {\bibfnamefont
  {H.~L.}\ \bibnamefont {Xin}}, \bibinfo {author} {\bibfnamefont
  {H.}~\bibnamefont {Li}}, \bibinfo {author} {\bibfnamefont {R.}~\bibnamefont
  {Li}}, \bibinfo {author} {\bibfnamefont {M.}~\bibnamefont {Fukuto}}, \bibinfo
  {author} {\bibfnamefont {K.~G.}\ \bibnamefont {Yager}}, \bibinfo {author}
  {\bibfnamefont {J.~S.}\ \bibnamefont {Kahn}}, \emph {et~al.},\ }\bibfield
  {title} {\bibinfo {title} {Ordered three-dimensional nanomaterials using
  dna-prescribed and valence-controlled material voxels},\ }\href@noop {}
  {\bibfield  {journal} {\bibinfo  {journal} {Nature materials}\ }\textbf
  {\bibinfo {volume} {19}},\ \bibinfo {pages} {789} (\bibinfo {year}
  {2020})}\BibitemShut {NoStop}%
\bibitem [{\citenamefont {Gerling}\ \emph {et~al.}(2015)\citenamefont
  {Gerling}, \citenamefont {Wagenbauer}, \citenamefont {Neuner},\ and\
  \citenamefont {Dietz}}]{gerling2015dynamic}%
  \BibitemOpen
  \bibfield  {author} {\bibinfo {author} {\bibfnamefont {T.}~\bibnamefont
  {Gerling}}, \bibinfo {author} {\bibfnamefont {K.~F.}\ \bibnamefont
  {Wagenbauer}}, \bibinfo {author} {\bibfnamefont {A.~M.}\ \bibnamefont
  {Neuner}},\ and\ \bibinfo {author} {\bibfnamefont {H.}~\bibnamefont
  {Dietz}},\ }\bibfield  {title} {\bibinfo {title} {Dynamic {DNA} devices and
  assemblies formed by shape-complementary, non--base pairing {3D}
  components},\ }\href@noop {} {\bibfield  {journal} {\bibinfo  {journal}
  {Science}\ }\textbf {\bibinfo {volume} {347}},\ \bibinfo {pages} {1446}
  (\bibinfo {year} {2015})}\BibitemShut {NoStop}%
\bibitem [{\citenamefont {Park}\ \emph {et~al.}(2019)\citenamefont {Park},
  \citenamefont {Park}, \citenamefont {Hur},\ and\ \citenamefont
  {Lee}}]{park2019design}%
  \BibitemOpen
  \bibfield  {author} {\bibinfo {author} {\bibfnamefont {S.~H.}\ \bibnamefont
  {Park}}, \bibinfo {author} {\bibfnamefont {H.}~\bibnamefont {Park}}, \bibinfo
  {author} {\bibfnamefont {K.}~\bibnamefont {Hur}},\ and\ \bibinfo {author}
  {\bibfnamefont {S.}~\bibnamefont {Lee}},\ }\bibfield  {title} {\bibinfo
  {title} {Design of dna origami diamond photonic crystals},\ }\href@noop {}
  {\bibfield  {journal} {\bibinfo  {journal} {ACS Applied Bio Materials}\
  }\textbf {\bibinfo {volume} {3}},\ \bibinfo {pages} {747} (\bibinfo {year}
  {2019})}\BibitemShut {NoStop}%
\bibitem [{\citenamefont {Rothemund}(2006)}]{Rothemund2006}%
  \BibitemOpen
  \bibfield  {author} {\bibinfo {author} {\bibfnamefont {P.~W.~K.}\
  \bibnamefont {Rothemund}},\ }\bibfield  {title} {\bibinfo {title} {{Folding
  DNA to create nanoscale shapes and patterns}},\ }\href
  {https://doi.org/10.1038/nature04586} {\bibfield  {journal} {\bibinfo
  {journal} {Nature}\ }\textbf {\bibinfo {volume} {440}},\ \bibinfo {pages}
  {297} (\bibinfo {year} {2006})}\BibitemShut {NoStop}%
\bibitem [{\citenamefont {Een}(2005)}]{een2005minisat}%
  \BibitemOpen
  \bibfield  {author} {\bibinfo {author} {\bibfnamefont {N.}~\bibnamefont
  {Een}},\ }\bibfield  {title} {\bibinfo {title} {Minisat: A sat solver with
  conflict-clause minimization},\ }in\ \href@noop {} {\emph {\bibinfo
  {booktitle} {Proc. SAT-05: 8th Int. Conf. on Theory and Applications of
  Satisfiability Testing}}}\ (\bibinfo {year} {2005})\ pp.\ \bibinfo {pages}
  {502--518}\BibitemShut {NoStop}%
\bibitem [{\citenamefont {Jun}\ \emph {et~al.}(2021)\citenamefont {Jun},
  \citenamefont {Wang}, \citenamefont {Parsons}, \citenamefont {Bricker},
  \citenamefont {John}, \citenamefont {Li}, \citenamefont {Jackson},
  \citenamefont {Chiu},\ and\ \citenamefont {Bathe}}]{athena}%
  \BibitemOpen
  \bibfield  {author} {\bibinfo {author} {\bibfnamefont {H.}~\bibnamefont
  {Jun}}, \bibinfo {author} {\bibfnamefont {X.}~\bibnamefont {Wang}}, \bibinfo
  {author} {\bibfnamefont {M.}~\bibnamefont {Parsons}}, \bibinfo {author}
  {\bibfnamefont {W.}~\bibnamefont {Bricker}}, \bibinfo {author} {\bibfnamefont
  {T.}~\bibnamefont {John}}, \bibinfo {author} {\bibfnamefont {S.}~\bibnamefont
  {Li}}, \bibinfo {author} {\bibfnamefont {S.}~\bibnamefont {Jackson}},
  \bibinfo {author} {\bibfnamefont {W.}~\bibnamefont {Chiu}},\ and\ \bibinfo
  {author} {\bibfnamefont {M.}~\bibnamefont {Bathe}},\ }\bibfield  {title}
  {\bibinfo {title} {{Rapid prototyping of arbitrary 2D and 3D wireframe DNA
  origami}},\ }\href@noop {} {\bibfield  {journal} {\bibinfo  {journal}
  {Nucleic Acids Research}\ }\textbf {\bibinfo {volume} {49}},\ \bibinfo
  {pages} {10265} (\bibinfo {year} {2021})}\BibitemShut {NoStop}%
\bibitem [{\citenamefont {Poppleton}\ \emph {et~al.}(2020)\citenamefont
  {Poppleton}, \citenamefont {Bohlin}, \citenamefont {Matthies}, \citenamefont
  {Sharma}, \citenamefont {Zhang},\ and\ \citenamefont {Šulc}}]{oxView}%
  \BibitemOpen
  \bibfield  {author} {\bibinfo {author} {\bibfnamefont {E.}~\bibnamefont
  {Poppleton}}, \bibinfo {author} {\bibfnamefont {J.}~\bibnamefont {Bohlin}},
  \bibinfo {author} {\bibfnamefont {M.}~\bibnamefont {Matthies}}, \bibinfo
  {author} {\bibfnamefont {S.}~\bibnamefont {Sharma}}, \bibinfo {author}
  {\bibfnamefont {F.}~\bibnamefont {Zhang}},\ and\ \bibinfo {author}
  {\bibfnamefont {P.}~\bibnamefont {Šulc}},\ }\bibfield  {title} {\bibinfo
  {title} {{Design, optimization and analysis of large DNA and RNA
  nanostructures through interactive visualization, editing and molecular
  simulation}},\ }\href@noop {} {\bibfield  {journal} {\bibinfo  {journal}
  {Nucleic Acids Research}\ }\textbf {\bibinfo {volume} {48}},\ \bibinfo
  {pages} {e72} (\bibinfo {year} {2020})}\BibitemShut {NoStop}%
\bibitem [{\citenamefont {Cock}\ \emph {et~al.}(2009)\citenamefont {Cock},
  \citenamefont {Antao}, \citenamefont {Chang}, \citenamefont {Chapman},
  \citenamefont {Cox}, \citenamefont {Dalke}, \citenamefont {Friedberg},
  \citenamefont {Hamelryck}, \citenamefont {Kauff}, \citenamefont
  {Wilczynski},\ and\ \citenamefont {de~Hoon}}]{biopython}%
  \BibitemOpen
  \bibfield  {author} {\bibinfo {author} {\bibfnamefont {P.~J.~A.}\
  \bibnamefont {Cock}}, \bibinfo {author} {\bibfnamefont {T.}~\bibnamefont
  {Antao}}, \bibinfo {author} {\bibfnamefont {J.~T.}\ \bibnamefont {Chang}},
  \bibinfo {author} {\bibfnamefont {B.~A.}\ \bibnamefont {Chapman}}, \bibinfo
  {author} {\bibfnamefont {C.~J.}\ \bibnamefont {Cox}}, \bibinfo {author}
  {\bibfnamefont {A.}~\bibnamefont {Dalke}}, \bibinfo {author} {\bibfnamefont
  {I.}~\bibnamefont {Friedberg}}, \bibinfo {author} {\bibfnamefont
  {T.}~\bibnamefont {Hamelryck}}, \bibinfo {author} {\bibfnamefont
  {F.}~\bibnamefont {Kauff}}, \bibinfo {author} {\bibfnamefont
  {B.}~\bibnamefont {Wilczynski}},\ and\ \bibinfo {author} {\bibfnamefont
  {M.~J.~L.}\ \bibnamefont {de~Hoon}},\ }\bibfield  {title} {\bibinfo {title}
  {{Biopython: freely available Python tools for computational molecular
  biology and bioinformatics}},\ }\href@noop {} {\bibfield  {journal} {\bibinfo
   {journal} {Bioinformatics}\ }\textbf {\bibinfo {volume} {25}},\ \bibinfo
  {pages} {1422} (\bibinfo {year} {2009})}\BibitemShut {NoStop}%
\bibitem [{\citenamefont {Bussi}\ \emph {et~al.}(2007)\citenamefont {Bussi},
  \citenamefont {Donadio},\ and\ \citenamefont
  {Parrinello}}]{bussi2007canonical}%
  \BibitemOpen
  \bibfield  {author} {\bibinfo {author} {\bibfnamefont {G.}~\bibnamefont
  {Bussi}}, \bibinfo {author} {\bibfnamefont {D.}~\bibnamefont {Donadio}},\
  and\ \bibinfo {author} {\bibfnamefont {M.}~\bibnamefont {Parrinello}},\
  }\bibfield  {title} {\bibinfo {title} {Canonical sampling through velocity
  rescaling},\ }\href@noop {} {\bibfield  {journal} {\bibinfo  {journal} {The
  Journal of chemical physics}\ }\textbf {\bibinfo {volume} {126}},\ \bibinfo
  {pages} {014101} (\bibinfo {year} {2007})}\BibitemShut {NoStop}%
\bibitem [{\citenamefont {Kern}\ and\ \citenamefont
  {Frenkel}(2003)}]{kern2003fluid}%
  \BibitemOpen
  \bibfield  {author} {\bibinfo {author} {\bibfnamefont {N.}~\bibnamefont
  {Kern}}\ and\ \bibinfo {author} {\bibfnamefont {D.}~\bibnamefont {Frenkel}},\
  }\bibfield  {title} {\bibinfo {title} {Fluid--fluid coexistence in colloidal
  systems with short-ranged strongly directional attraction},\ }\href@noop {}
  {\bibfield  {journal} {\bibinfo  {journal} {The Journal of chemical physics}\
  }\textbf {\bibinfo {volume} {118}},\ \bibinfo {pages} {9882} (\bibinfo {year}
  {2003})}\BibitemShut {NoStop}%
\bibitem [{\citenamefont {Weeks}\ \emph {et~al.}(1971)\citenamefont {Weeks},
  \citenamefont {Chandler},\ and\ \citenamefont
  {Andersen}}]{weeks::jcp::54::1971}%
  \BibitemOpen
  \bibfield  {author} {\bibinfo {author} {\bibfnamefont {J.~D.}\ \bibnamefont
  {Weeks}}, \bibinfo {author} {\bibfnamefont {D.}~\bibnamefont {Chandler}},\
  and\ \bibinfo {author} {\bibfnamefont {H.~C.}\ \bibnamefont {Andersen}},\
  }\bibfield  {title} {\bibinfo {title} {Role of repulsive forces in
  determining the equilibrium structure of simple liquids},\ }\href@noop {}
  {\bibfield  {journal} {\bibinfo  {journal} {Journal of Chemical Physics}\
  }\textbf {\bibinfo {volume} {54}},\ \bibinfo {pages} {5237} (\bibinfo {year}
  {1971})}\BibitemShut {NoStop}%
\end{thebibliography}%

\onecolumngrid
\appendix
\newpage

\section{SUPPLEMENTARY INFORMATION}

\setcounter{figure}{0}
 \makeatletter 
 \renewcommand{\thefigure}{S\@arabic\c@figure}
 \setcounter{equation}{0}
 \renewcommand{\theequation}{S\@arabic\c@equation}
 \setcounter{table}{0}
 \renewcommand{\thetable}{S\@arabic\c@table}
  \setcounter{section}{0}
 \renewcommand{\thesection}{S\@Roman\c@section}

 \pagenumbering{gobble} 

\subsection{SI SAT-assembly clauses}

\begin{table}[h!]
	\centering
	\begin{tabular}{c|c|c} 
		\hline
		Id  & Clauses & Boolean expression \\ 
		\hline
		(i) & $C^{\rm int}_{c_i,c_j,c_k}$ & $ \neg x^{\rm int}_{c_i,c_j} \lor \neg x^{\rm int}_{c_i,c_k}$  \\
		(ii) & $C^{\rm pcol}_{s,p,c_k,c_l}$ & $\neg x^{pcol}_{s,p,c_k} \lor \neg x^{\rm pcol}_{s,p,c_l}$  \\ 
		(iii) & $ C^L_{l,s_i,o_i,s_j,o_j}$ & $\neg x^L_{l,s_i,o_i} \lor \neg x^L_{l,s_j,o_j}$ \\ 
		 (iv) & $C^{\rm lint}_{l_i,k_i,l_j,k_j,c_i,c_j}$ &  $(x^A_{l_i,k_i,c_i} \land  x^A_{l_j,k_j,c_j}) \implies x^{\rm int}_{c_i,c_j}$ \\
		(v) & $C^{\rm LS}_{l,s,o,c,k}$ &  $ x^L_{l,s,o} \implies \left( x^A_{l, k, c} \iff x^{\rm pcol}_{s, \phi_o(k), c} \right) $ \\
		(vi) & $C^{\rm all\,s.}_{s}$ & $  \bigvee_{\forall l , o } x^L_{l,s,o} $ \\
	    (vii)& $C^{\rm all\,c.}_{c}$ & $ \bigvee_{\forall s , p  } x^{\rm pcol}_{s,p,c}$ \\
		\hline
	\end{tabular}
	\caption{SAT clauses and variables.
	}
	\label{table:sat}
\end{table}

The color interaction is given by binary variables $x^{\rm int}_{c_i,c_j}$ which are 1 if color $c_i$ is compatible with color $c_j$ and 0
	otherwise. The patch coloring for each PP species is described by binary variables $x^{\rm pcol}_{s,p,c}$ which are 1 if patch $p$ of species $s$ has color $c$ and 0 otherwise. 
	The arrangement of the particle species in the lattice is described by $x^{L}_{l,s,o}$ which is 1 if the position $l$ is occupied by a PP of species $s$ in the specific orientation $o$. The mapping $\phi_o(k) = p$ for a given orientation $o$ means that PP's patch $p$ overlaps with slot $k$ in a given lattice position. The variable $x^A_{l,k,c}$ is 1 if slot $k$ of lattice position $l$ is occupied by a patch with color $c$ and 0 otherwise. The clauses and variables are defined for all possible combinations of colors $c \in [1,N_c]$, patches $p \in [1,N_p]$, slots $k \in [1,N_p]$,  PP species $s \in [1,N_s]$, orientations $o \in [1,N_o]$, and lattice positions $l \in [1,L]$. Clauses $C^{\rm lint}$ are defined only for slots $k_i,k_j$ that are in contact in neighboring lattice positions $l_i, l_j$. 
	 For a given $s$, clause $C^{\rm all\,s.}_{s}$ is defined as a list of $x^L_{l,s,o}$ for all possible values of $l$ and $o$, joined by disjunctions. Clause $C^{\rm all\,c.}_{c}$ is defined analogously.
	 The SAT problem is a conjunction of all clauses. The color interaction matrix is given by the indices of the $x^{\rm int}_{c_i,c_j}$ that are true in the found solution, and the coloring (which particle species is assigned which color) is given by the indices of variables $x^{\rm pcol}_{s,p,c}$  that are true.
	 We first define the SAT problem using the topology of CD lattice as given in Table \ref{tab:doublediamond}. For each identified solution, we formulate a new SAT problem
	 with additional new clauses that require $x^{\rm int}_{c_i,c_j}$ and  $x^{\rm pcol}_{s,p,c}$ variables that correspond to the identified solution are true. We then reformulate
	 clauses $C^L$ to correspond to the HD topology (Table \ref{tab:megahexagonal}). If such a SAT problem is satisfiable, we discard the solution and repeat the process, until we find a 
	 solution that satisifies CD topology but does not satisify the HD topology. 
	 
\subsection{SII Cubic and Hexagonal diamond topologies}

We include here the topology for a unit cell of cubic diamond crystal lattice 16-unit cell, and for hexagonal diamond lattice 32-unit cell. These cells are created by merging smaller unit cells of cubic diamond and hexagonal diamond respectively.
 \begin{table}
 \centering
 \begin{tabular}{c|c|c|c}
 \hline
 Position $l_i$ & Slot $s_i$ & Position $l_j$ & Slot $s_j$\\ \hline
2 & 4 & 15 & 4  \\
10 & 2 & 14 & 4  \\
4 & 1 & 8 & 3  \\
12 & 3 & 13 & 2  \\
3 & 2 & 6 & 3  \\
1 & 4 & 5 & 4  \\
11 & 4 & 16 & 4  \\
2 & 3 & 5 & 3  \\
4 & 4 & 7 & 2  \\
7 & 4 & 10 & 4  \\
12 & 2 & 14 & 2  \\
3 & 3 & 7 & 3  \\
11 & 1 & 13 & 1  \\
2 & 2 & 6 & 4  \\
4 & 3 & 5 & 2  \\
1 & 1 & 16 & 1  \\
9 & 3 & 14 & 1  \\
8 & 2 & 10 & 1  \\
7 & 1 & 9 & 2  \\
12 & 1 & 16 & 3  \\
3 & 4 & 8 & 4  \\
11 & 2 & 14 & 3  \\
2 & 1 & 16 & 2  \\
1 & 2 & 15 & 1  \\
9 & 4 & 13 & 4  \\
8 & 1 & 9 & 1  \\
10 & 3 & 13 & 3  \\
4 & 2 & 6 & 2  \\
12 & 4 & 15 & 2  \\
3 & 1 & 5 & 1  \\
1 & 3 & 6 & 1  \\
11 & 3 & 15 & 3  \\
  \hline
  \end{tabular}
 \caption{\label{tab:doublediamond} Cubic diamond 16-unit cell topology: List of lattice positions $l_i$ and $l_j$ that are neighbors in the unit cell of the lattice and their respective slot numbers $s_i$, $s_j$ through which they are bound. The unit cell of size 16 is obtained by pasting together two unit cells of size 8.}
 \end{table}

 \begin{table}
 \small
 \centering
 \begin{tabular}{c|c|c|c}
 \hline
 Position $l_i$ & Slot $s_i$ & Position $l_j$ & Slot $s_j$\\ \hline
8 & 4 & 17 & 1  \\
2 & 4 & 23 & 1  \\
10 & 2 & 14 & 2  \\
4 & 1 & 6 & 2  \\
12 & 3 & 29 & 1  \\
22 & 3 & 26 & 2  \\
17 & 3 & 29 & 2  \\
3 & 2 & 7 & 2  \\
27 & 3 & 30 & 4  \\
7 & 3 & 11 & 1  \\
18 & 3 & 30 & 2  \\
1 & 4 & 24 & 1  \\
11 & 4 & 15 & 4  \\
28 & 3 & 29 & 4  \\
8 & 3 & 12 & 1  \\
2 & 3 & 14 & 1  \\
13 & 3 & 28 & 1  \\
23 & 3 & 27 & 2  \\
4 & 4 & 22 & 1  \\
9 & 2 & 13 & 2  \\
17 & 4 & 24 & 2  \\
27 & 4 & 31 & 4  \\
24 & 3 & 28 & 2  \\
7 & 4 & 18 & 1  \\
12 & 2 & 13 & 4  \\
19 & 2 & 21 & 4  \\
26 & 4 & 32 & 3  \\
3 & 3 & 15 & 1  \\
2 & 2 & 7 & 1  \\
20 & 4 & 24 & 4  \\
4 & 3 & 16 & 1  \\
1 & 1 & 5 & 1  \\
14 & 3 & 27 & 1  \\
9 & 3 & 31 & 1  \\
19 & 3 & 31 & 2  \\
10 & 4 & 16 & 2  \\
18 & 2 & 22 & 2  \\
20 & 3 & 32 & 2  \\
3 & 4 & 21 & 1  \\
25 & 3 & 29 & 3  \\
11 & 2 & 14 & 4  \\
5 & 3 & 9 & 1  \\
2 & 1 & 6 & 1  \\
15 & 3 & 25 & 1  \\
6 & 4 & 20 & 1  \\
1 & 2 & 8 & 1  \\
9 & 4 & 15 & 2  \\
19 & 4 & 23 & 4  \\
16 & 3 & 26 & 1  \\
10 & 3 & 32 & 1  \\
21 & 3 & 25 & 2  \\
4 & 2 & 8 & 2  \\
12 & 4 & 16 & 4  \\
28 & 4 & 32 & 4  \\
17 & 2 & 21 & 2  \\
3 & 1 & 5 & 2  \\
25 & 4 & 31 & 3  \\
18 & 4 & 23 & 2  \\
5 & 4 & 19 & 1  \\
26 & 3 & 30 & 3  \\
20 & 2 & 22 & 4  \\
6 & 3 & 10 & 1  \\
1 & 3 & 13 & 1  \\
11 & 3 & 30 & 1  \\
  \hline
  \end{tabular}
 \caption{\label{tab:megahexagonal} Hexagonal 32-unit cell topology, obtained by joining 4 unit cell lattices. List of lattice positions $l_i$ and $l_j$ that are neighbors in the unit cell and their respective slot numbers $s_i$, $s_j$ through which they are bound }
 \end{table}

 \begin{table}
 \centering
 \begin{tabular}{c|c}
 \hline
 Orientation $o$ & Mapping $\phi_o$ \\ \hline
 1 & (1,2,3,4) \\ 
 2 & (1,4,2,3) \\
3 & (1,3,4,2) \\
4 & (2,4,3,1) \\
5 & (2,1,4,3) \\
6 & (2,3,1,4) \\
7 & (4,1,3,2) \\
8 & (4,2,1,3) \\
9 & (4,3,2,1) \\
10 & (3,1,2,4) \\
11 & (3,4,1,2) \\
12 & (3,2,4,1) \\
  \hline
  \end{tabular}
 \caption{\label{tab:rotation} List of orientations $o$ for a PP with a tetrahedral symmetry of patch positions. Each orientation determines a mapping $\phi_o$, which specifies which patches overlap with patches in the original position before applying the rotation. The first mapping for $o = 1$ corresponds to no rotation.}
 \end{table}

\end{document}